\documentclass[12pt]{article}

\usepackage[english]{babel}

\usepackage[letterpaper,top=2cm,bottom=2cm,left=3cm,right=3cm,marginparwidth=1.75cm]{geometry}

\usepackage[utf8]{inputenc} % allow utf-8 input
\usepackage[T1]{fontenc}    % use 8-bit T1 fonts
\usepackage{hyperref}       % hyperlinks
\usepackage{url}            % simple URL typesetting
\usepackage{booktabs}       % professional-quality tables
\usepackage[flushleft]{threeparttable}
\usepackage{amsfonts}       % blackboard math symbols
\usepackage{nicefrac}       % compact symbols for 1/2, etc.
\usepackage{microtype}      % microtypography
\usepackage{amsthm,amsmath,amsfonts, amssymb}
\usepackage{stmaryrd}
\usepackage{multirow}
\usepackage{graphicx}
\usepackage{xcolor}
\usepackage{mathrsfs,mathtools}
\usepackage{bm, bbm}
\usepackage{authblk}
\usepackage[numbers]{natbib}
\usepackage[noend, algoruled, linesnumbered]{algorithm2e}
\SetKwInOut{Input}{Input}
\usepackage{pdflscape}
\usepackage{threeparttable}
\usepackage{arydshln}
\usepackage{float}
\usepackage{float}

\usepackage{caption}
\usepackage{subcaption}
\usepackage{wrapfig}

\RestyleAlgo{ruled}
\newlength\myindent
\newcommand{\cmtH}[1]{\textcolor{black}{#1}}

\newtheorem{proposition}{Proposition}
\newtheorem{remark}{Remark}
\newtheorem{lemma}{Lemma}
\newtheorem{definition}{Definition}

\newcommand*{\ie}{\emph{i.e.}{}}

\newcommand{\indep}{\perp \!\!\! \perp}

\title{\bf 
New User Event Prediction\\Through the Lens of Causal Inference}

\author[1]{Henry Shaowu Yuchi}
\affil[1]{Los Alamos National Laboratory}
\author[2]{Shixiang Zhu}
\affil[2]{Carnegie Mellon University}
\author[3]{\authorcr Li Dong}
\author[3]{Yigit M. Arisoy}
\author[3]{Matthew C. Spencer}
\affil[3]{Amazon}

\date{}

\begin{document}
\maketitle

\begin{abstract}
Modeling and analysis for event series generated by users of heterogeneous behavioral patterns are closely involved in our daily lives, including credit card fraud detection, online platform user recommendation, and social network analysis. 
The most commonly adopted approach to this task is to assign users to behavior-based categories and analyze each of them separately.
However, this requires extensive data to fully understand the user behavior, presenting challenges in modeling newcomers without significant historical knowledge.
In this work, we propose a novel discrete event prediction framework for new users with limited history, without needing to know the user's category. We treat the user event history as the ``treatment'' for future events and the user category as the key confounder. Thus, the prediction problem can be framed as counterfactual outcome estimation, where each event is re-weighted by its inverse propensity score.
% with the new user model trained on an adjusted dataset where each event is re-weighted by its inverse propensity score.
% becomes one of counterfactual outcome estimation, with the new user model trained on an adjusted dataset where each event is reweighted by its inverse propensity score. 
We demonstrate the improved performance of the proposed framework with a numerical simulation study and two real-world applications, including Netflix rating prediction and seller contact prediction for customer support at Amazon.
\end{abstract}

\section{Introduction}

Sequential user records data have become ubiquitously mired in everyday life. These records of event data typically contain several pieces of information, including when the events take place (\ie, \textit{temporal}) and where they do so (\ie, \textit{spatial}), in addition to others. The temporal point process has already proven to be a powerful tool in modeling such sequential spatio-temporal data with a wide range of applications, including earthquake prediction \citep{ogata1998space,schoenberg2003multidimensional}, crime modeling  \citep{mohler2014marked,mohler2011self,zhu2022spatiotemporal}, credit card fraud detection \citep{zhu2020adversarial}, infectious disease forecasting \citep{chiang2022hawkes,garetto2021time}, and social network studies \citep{fox2016modeling,wong2006spatial}. Especially, temporal point process models have been widely applied for user event prediction in e-commerce \citep{ge2021towards, li2021user} given history, as the online marketplace is rapidly gaining popularity over recent years. 

\cmtH{Nowadays, platforms with a large number of users typically categorize them to improve efficiency and service quality, which is often based on their behavioral profile in practice \citep{pereira2023spotify}. This is conducted based on a wide range of user characteristics including their background information and business models, beyond simply their event history. For instance, online e-commerce platforms typically put sellers into different categories based on their business sector, product offerings, and inventory needs. This extensive process is usually both time-consuming and costly.
% This process usually has to depend on extensive manual processes such as interviews and business profiling by platform representatives, which is both time-consuming and costly \citep{gomez2022automatic}.
Consequently, many newer users do not yet have their category information established.}

Predicting events for new users given their limited history presents two key challenges: 
($i$) It demands a deep understanding of user characteristics and historical behavior to identify user patterns. This task becomes particularly difficult when the user base is large, as categorizing new users accurately is often infeasible due to their limited history data.
($ii$) The distribution of future events across users may be imbalanced, making it impractical to apply a uniform model to both established and new users, as such a model would tend to favor those with more extensive event histories.
 
% It requires extensive knowledge of user characteristics and historical events to understand user patterns. This is necessary to ascertain user categorization when the user base is large, which is often impossible for newer users. 

% The heterogeneous behaviors of users make it impractical to apply a single uniform model to both new and existing users, leading to significant performance discrepancies between different users. In practice, experienced professionals categorize users to improve service efficiency and quality. This categorization involves collecting background information, past behavior, user context, and more, and it is a time-consuming manual process. Consequently, new users often lack category information, which leads to several issues: 
% ($i$) The absence of category information causes confounding bias to the event prediction due to the complex dependencies between the future behavior, historical events, and user categories, resulting in modeling distortions. ($ii$) The distribution of user events across categories is imbalanced. Without category information, modeling user events tends to favor categories with more users or more data.

\begin{figure}
% \vspace{-.15in}
\centering
    \includegraphics[width=.8\linewidth]{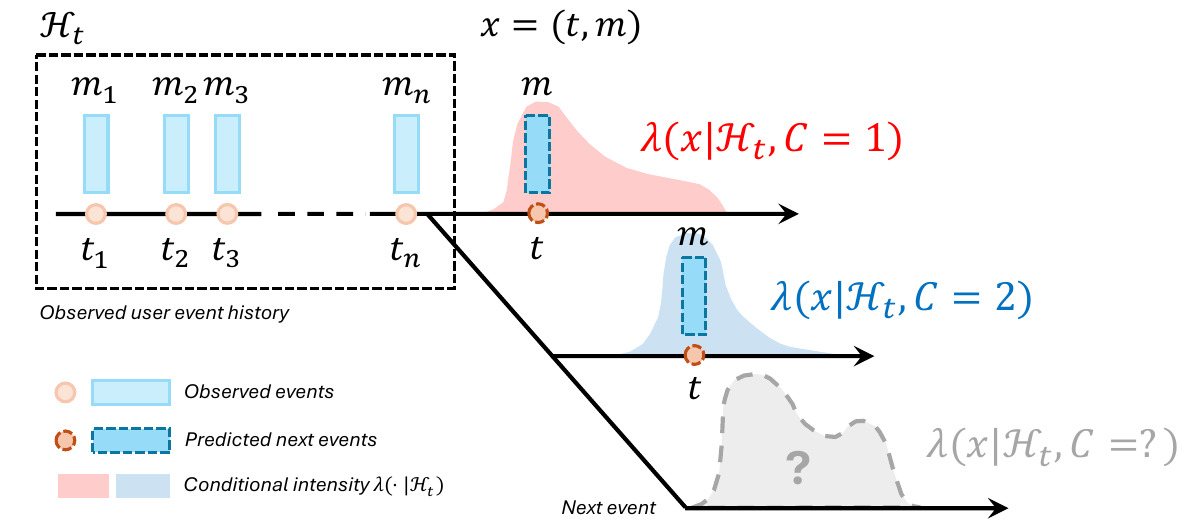}
    \captionof{figure}{The goal is to predict the next event for a new user without knowing its category. The distribution of the next event is influenced by its category even with the same event history. Each event $x$ consists of its occurrence time $t$ and associated mark data $m$. The conditional intensity $\lambda(x|\mathcal{H}_t, C)$ represents the occurrence rate of the next event $x$ given its history $\mathcal{H}_t$ and the user's category $C$.}
    \label{fig:convergence_cluster}
%\vspace{-.15in}
\end{figure}

To address these challenges, we formulate a \emph{new user event prediction} problem, as illustrated in Fig.~\ref{fig:convergence_cluster}.
We investigate the problem by asking a ``what if'' question, \ie, ``When would the next event occur and what would be the mark of the next event given this new user had engaged in a series of activities, without knowing the user's category?'' 
We answer this question by \emph{viewing a user's event history as a form of ``intervention'' to its next event and its feature as the ``confounder''}. 
\cmtH{The objective is to study the effect of certain history across the entire user base on users' future behavior, irrespective of each user's specific category.}
This is achieved through reweighting the observed user event sequences with feature information. 
%The weights are motivated by the inverse propensity score from causal inference, which is the inverse probability of a user having a specific history record given their observed characteristics. 
% This is equivalent to asking that ``Provided what is known about the user, how likely they will be given the corresponding event history (treatment)''. 
Inverse propensity scoring is utilized to balance users based on their propensity scores, giving more weight to those underrepresented otherwise in the model. This makes event prediction unbiased given the heterogeneous behavior of multiple users. 
Here, the propensity scores are estimated from the intervention/treatment variables, while the latter are updated as the model is fit with inputs including the propensity scores. 
% Therefore, there exists a structure of \textit{cyclic dependence} between the weights and model parameters which requires more attention to handle.

The main contributions of this paper are threefold: ($i$) This work enables event prediction for new users without the necessity for their user feature information, which is hard to acquire. ($ii$) This work aims to achieve unbiased user event prediction through a novel counterfactual framework by viewing the user's history as the intervention for its future event. It proposes and applies a new alternating learning algorithm between updating the inverse propensity scores and the model estimation. ($iii$) It is demonstrated that the proposed framework can achieve superior performance with both simulation studies and real applications, validating the potential of the method to be more widely implemented.

% \begin{enumerate}
%     \item This work enables event prediction for new users without the necessity for their user feature information which is hard to acquire.
%     \item This work aims to achieve unbiased user event prediction through a novel counterfactual framework by viewing the user's history as the intervention to its future event. It proposes and applies a new alternating learning algorithm between updating the inverse propensity scores and the model estimation.
%     \item It is demonstrated in this paper that the proposed framework can achieve superior performance with both simulation studies and real applications, validating the potential of the method to be more widely implemented.
% \end{enumerate}

\paragraph{Related Works}
% \label{sec:lit_review}
Point process models have been a powerful and popular tool used for discrete event prediction. 
The seminal work by \cite{hawkes1971spectra} introduced the concept of the self-exciting temporal Hawkes point process.
% , where an event, defined by its time, is likely to trigger another determined by the conditional intensity via triggering kernel functions. 
A similar self-correcting point process was introduced in \cite{ogata1984inference}. Spatio-temporal point process models extend the event space of observations to include location \citep{daley2003introduction}. 
%and features beyond time and location can be included in the model as additional mark variables \cite{daley2003introduction}. 
% With that, point process models are capable of tackling complex event sequence data, often with much more information beyond temporal and spatial components. 
Point process models have been thoroughly reviewed in \cite{daley2003introduction,last2017lectures,reinhart2018review}.
% Point process models have been comprehensively reviewed in \cite{last2017lectures} for its mathematical foundation and in \cite{reinhart2018review,daley2003introduction} for its structure and deployment.
There have been attempts at event prediction for multiple users given past sequences. The work by \cite{du2015time} establishes a user-item-event framework utilizing a self-exciting Hawkes point process model and exploits the low-rank property of the data. 
%There have been previous attempts at predicting upcoming events for multiple users given their past sequences. The work by \cite{du2015time} establishes a user-item-event framework utilizing the point process model.
%It models the behavior of each user with a self-exciting Hawkes point process model, and then the authors exploit the low-rank property of the framework to construct an efficient way of modeling multiple users at the same time.
However, in the problem pursued by this work, users' behavior is highly heterogeneous when there are many of them, which no longer satisfies the low-rank assumption.

With the rise to prominence of neural networks (NN) that can model intricate and complex relationships between events, it has been taken up to further enhance the expressiveness of point process models. We introduce two such major research efforts. 
The first one is the Neural Hawkes process (NH) proposed by \cite{mei2017neural}, which is a self-modulating multivariate point process. It models the the conditional intensity function using a neural network constructed with a high-dimensional hidden state variable which in turn depends on another set of memory variables in continuous-time Long Short Term Memory (LSTM) \citep{graves2012supervised}. 
% The multivariate formulation then enables the model to encompass event sequences contributed by multiple sources, making the model more versatile. 
The other work is the Recurrent Marked Temporal Point Process (RMTPP) proposed in \cite{du2016recurrent}, where the event history is encoded using a hidden embedding layer pushed through a recurrent neural network (RNN). 
% In the model structure, the event time and mark components are fed to the RNN where the hidden embedding learns a nonlinear dependency over both components from past events. 
In both works, the utilization of the hidden history layer in NN facilitates the encapsulation of past event information. Other similar works include modeling the conditional intensity also using RNN \citep{zhang2020self} and reinforcement learning \citep{li2018learning}. 

Our work is inspired by the marginal structural models in causal inference \citep{robins1999association,robins1999estimation,robins2000marginal}, among other classical approaches to time-varying treatment effect estimation including the g-formula and structural nested models \citep{robins1986new, robins1994correcting,rubin1978bayesian}. We regard the history embedding as the treatment/intervention variable for estimation, which can be continuous \citep{zou2022counterfactual} or of high dimension \citep{zou2020counterfactual}. 
Recently new efforts have been spotted in this subject, estimating treatment effect by spline and kernel regression \citep{kennedy2017non,nie2021vcnet}, Gaussian process \citep{chen2023multi,schulam2017reliable}, RNN-based models \citep{li2021g,lim2018forecasting}, adversarial networks \citep{bica2020estimating,yoon2018ganite}, or by transformer models \citep{melnychuk2022causal}. A generative model producing counterfactual samples is proposed as well \citep{wu2023counterfactual}. 

New user behavior has been studied in collaborative filtering \citep{ahn2008new, bobadilla2012collaborative}. It is often treated as a cold start problem and the performance from different recommender systems vary \citep{kluver2014evaluating}. Otherwise there is sparse literature on event prediction for new users which we focus on in this work.

\section{Preliminaries}
\label{sec:prelim}
\subsection{Marked Temporal Point Processes}

Marked temporal point processes (MTPPs) \citep{reinhart2018review} consist of a sequence of \emph{discrete events} over time. Each event is associated with a {\it mark} that contains detailed information about the event, such as user intent in our case study. 
   Let $T > 0$ be a fixed time-horizon, and $\mathscr{M} \subseteq \mathbb{R}^d$ be the space of marks. We denote the space of observation by $\mathscr{X} = [0, T) \times \mathscr{M}$ and the $i$th data point in the discrete event sequence by
\begin{equation*}
    x_i = (t_i, m_i), \quad t_i \in [0, T) , \quad m_i \in \mathscr M,
\end{equation*}
where $t_i$ is the event time and $m_i$ represents the corresponding mark. 
Let $N_t$ be the number of events up to time $t < T$ (which is random).
We define $\mathcal{H}_t := \{x_1 , x_2 , \dots, x_{N_t}\}$, denoting historical events.
Let $\mathbb{N}$ be the counting measure on $\mathscr{X}$, \ie, for any measurable $S \subseteq \mathscr{X}$, $\mathbb{N}(S) = |\mathcal{H}_t \cap S|.$ 
For any function $g:\mathscr{X}\to\mathbb{R}$, the integral with respect to the counting measure is defined as $\int_{S} g(x) d\mathbb N(x) = \sum_{x_i\in\mathcal H_T\cap S} g(x_i)$.

The distribution of events in MTPPs is typically characterized via the conditional intensity function $\lambda$, which is defined to be the occurrence rate of events in the marked temporal space $\mathscr{X}$ given the events' history $\mathcal{H}_{t(x)}$, \ie,
\begin{equation}
    \lambda(x| \mathcal{H}_{t(x)} ) = \mathbb{E}\left( d\mathbb{N}(x) | \mathcal{H}_{t(x)} \right) / dx,
    \label{eq:cond-intensity}
\end{equation}
where $t(x)$ extracts the occurrence time of the possible event $x$. Given the conditional intensity function $\lambda$, the corresponding conditional probability density function (PDF) can be expressed as
\begin{align}
\begin{split}
    f(x| \mathcal{H}_{t(x)} ) =  \lambda(x| \mathcal{H}_{t(x)} ) 
     \cdot \exp \left( -\int_{[t_n, t(x)) \times \mathscr{M}} \lambda(u| \mathcal{H}_{t(u)} ) du \right),
    \label{eq:cond-prob}
\end{split}
\end{align}
where $t_n$ denotes the time of the most recent event that occurred before time $t(x)$. 

The point process models can be fitted using maximum likelihood estimation (MLE). The log-likelihood of observing a sequence with $N_T$ events can therefore be obtained by
\begin{align}
\begin{split}
    \ell(x_1, \dots, x_{N_T}) = \int_{\mathscr X}  \log \lambda(x| \mathcal{H}_{t(x)} ) d \mathbb N(x) 
    -  \int_{\mathscr X}  \lambda(x| \mathcal{H}_{t(x)} ) dx.
\end{split}
\label{eq:log-likelihood}
\end{align}
See the derivations in Appendix~\ref{append:derivation-cond-prob}.
%\vspace{-0.1in}
\subsection{Neural Point Processes}
Neural point processes (NPPs) \citep{du2016recurrent, mei2017neural, zuo2020transformer} are commonly used for modeling complex event sequences. 
In NPPs, the event history is summarized using a \emph{history encoder}, often a recursive structure such as recurrent neural networks (RNNs) \citep{du2016recurrent, mei2017neural} or Transformers \citep{pmlr-v130-zhu21b, zuo2020transformer}, taking the history of event $x$ as its input and generating a low-dimensional and compact \emph{history embedding}, denoted by $h(x) \in \mathscr{H} \subseteq \mathbb{R}^q$. 
This history embedding represents an updated summary of the past events including $x$.
Assume there are $n$ observed events for a user, the history embedding of the next event $x$, $h(x)$, can be written as follows:
\begin{equation}
    h(x) = \phi(t(x) - t(x_n), m(x), h(x_n)),
    % h_i = \phi(\Delta t, h_{i-1}),
    \label{eq:history-var}
\end{equation}
where $\phi$ is a non-linear mapping, and $n$ denotes the index of the last observed event before $t(x)$. 
Therefore, the conditional intensity in \eqref{eq:cond-intensity} can be approximated by
\begin{equation}
    \lambda( x | \mathcal{H}_{t(x)}) \approx \lambda_\theta(x |h(x)),
    \label{eq:nn-lam} 
\end{equation}
where $\theta$ denotes the parameters of the model, including the parameters in $\lambda$ and the nonlinear mapping $\phi$. The corresponding conditional PDF can then be denoted by $f_\theta$ following \eqref{eq:cond-prob}. 
% This work looks into how this history embedding changes with respect to time, and attempts to model this change using its Markov properties.

\section{New User Event Prediction}
\label{sec:method}

The objective is to find an approach that enables unbiased prediction of the next immediate event $\widehat{x}_{n+1}$ for a new user based on their history $h(x)$ without the knowledge of the particular new user's category information $C$. \cmtH{In this section, we relay the category-agnostic intensity framework and the IPTW reweighting scheme. We also propose an alternate learning algorithm to fit the framework.}

\begin{figure}
% \vspace{-0.15in}
\centering
        \includegraphics[width=0.6\linewidth]{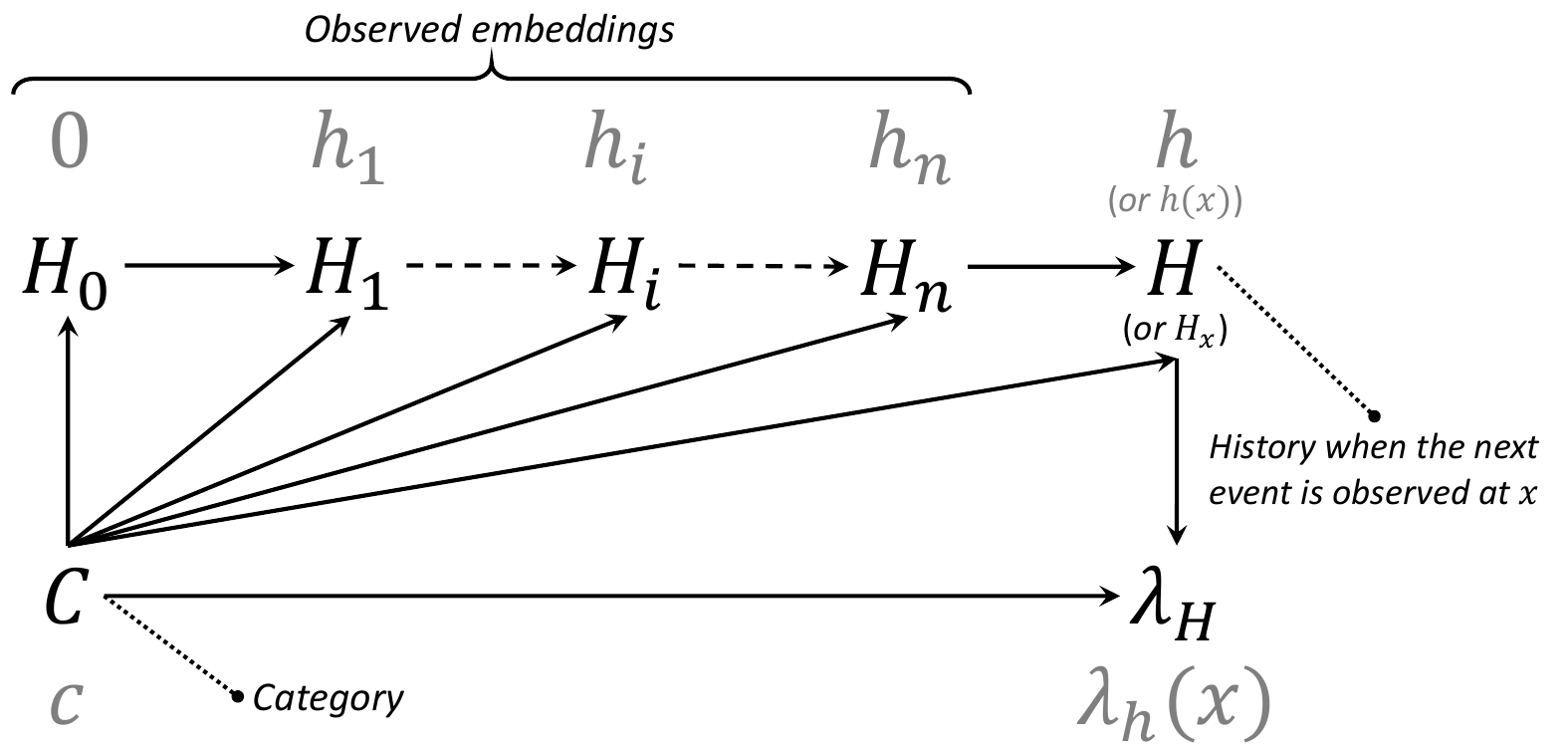}
        \captionof{figure}{Causal DAG between user category, history variables, and category-agnostic intensity. The lowercase notations represent the realizations of these variables. We use $n$ to denote the number of observed events in the history and $h(x)$ (or simply $h$) denotes the history when we observe the next ($n+1$)-th event at $x$.}
        \label{fig:causal-dag}
% \vspace{-0.15in}
\end{figure}

\subsection{Category-agnostic Intensity Estimation}

By viewing the event history $h(x)$ as the intervention and the user category $C$ as the confounder, our goal is to provide an unbiased estimation of the effect of event history $h(x)$ on the next event $x$, while accounting for the influence of the confounding variable $C$. 
% We represent $h(x)$ and $H_x$ using $h$ and $H$, respectively.
We first introduce a new notion that treats the history embedding as a random variable, termed \emph{random history variable}:
\begin{definition}[Random history variable]
    The history embedding $h_i$ and $h(x)$ are considered realizations of two random variables, denoted by $H_i$ and $H_{x}$, respectively, and collectively referred to as history variables. We note that $H_{x}$ depends on $H_n$ and $H_i$ depends on $H_{i-1}$ according to \eqref{eq:history-var}.
\end{definition}
%\vspace{-0.05in}
Further, we show that history variables possess the Markov property \citep{kuo2018markov}:
\begin{lemma}[Markov property]
The history embedding $h_0, h_1,\dots, h_i$, and $h(x)$ possess the Markov property, formulated as follows:
\begin{equation}
\lambda(x|h(x)) = \lambda(x|h(x), h_n, \dots, h_i,\dots, h_1, h_0).
\label{eq:markov}
\end{equation}
This can be shown from the definition of $h(x)$ in \eqref{eq:history-var}, where $h(x)$ can be specified by $x$ and $h_n$.
\label{lemma:markov}
\end{lemma}
% In the rest of the paper, we represent $h(x)$ and $H_x$ using $h$ and $H$, respectively.%, for notional simplicity.
% \woody{I put them here temporarily...}
%\vspace{-0.05in}
For more details on the Markov property, see Appendix~\ref{sec:markov}. 
Our proposed approach is grounded in Lemma~\ref{lemma:markov} that the history variable of a user is influenced by both the previous history and its user category, leading to the following definition:
\begin{definition}[Dependence of users' future event]
    The immediate next event of a user is not only affected by its up-to-date history $H_{x}$ reflected in \eqref{eq:nn-lam}, but also by the user's category $C$. 
    The causal directed acyclic graph (DAG) of these variables is shown in Fig.~\ref{fig:causal-dag}.
\end{definition}
%\vspace{-0.05in}
% \cmtH{Assume the occurrence rate of the next event $x$, given a user has already taken the first $n$ activities represented by $H=h$ and its category, can be defined by
%     \[
%     \lambda(x|h, c) = \mathbb{E}\left( d \mathbb{N}(x|h, c) \right)/ dx,
%     \]
% where $\mathbb{N}(x|h,c)$ is the counting measure of a user if the user ``receives'' history $h$ while its category is $c$.
% % Additionally, assume that for some certain $h \in \mathscr{H}$ and $c \in \mathscr{C}$, 
% % \begin{equation}
% % \mathbb{E}_Y\left[ \lambda(x|H=h, C)\right] \indep H | C=c.
% % \label{eq:ignor}
% % \end{equation}
% }
Given that, we can refer to our estimand as \emph{category-agnostic intensity}, which is formally defined as follows:
\begin{definition}[Category-agnostic intensity]
    The category-agnostic occurrence rate of the next event $x$, given a user has already taken the first $n$ activities represented by $H=h$, can be defined by 
    \[
    \lambda_{h}(x) = \mathbb{E}[d \mathbb{N}_{h}(x)] / dx,
    \]
    where $\mathbb{N}_{h}(x)$ is the counting measure of a user if the user ``receives'' history $h$. 
    \label{def:counterfactual}
\end{definition}
% \cmtH{Following this, the conditional intensity can be similarly defined by 
% \[
% \lambda(x|h, c) = \mathbb{E}\left( d \mathbb{N}(x|h, c) \right)/ dx,
% \]
% where $\mathbb{N}(x|h,c)$ is the counting measure of a user if the user ``receives'' history $h$ while its category is $c$.}

%\vspace{-0.05in}
We note that the category-agnostic intensity $\lambda_{h}(x)$ describes the interventional distribution of the next event, assuming the same event history across all users. This is notably different from the un-adjusted conditional intensity $\lambda(x|h)$, which does not factor in 
the disparity of users' category across the entire population. 
% the difference of users' category.
To identify the category-agnostic intensity, we state the following three key assumptions:
\begin{enumerate}
    \item \emph{Consistency}: Provided the history is $h$, then $\lambda_h(x)$ is the potential intensity under history $h$. Formally, $H=h \Rightarrow \lambda(x|H=h) = \lambda_h(x)$.
    \item \emph{Positivity}: If $\mathbb{P}(H_i = h_i, H_{i-1} = h_{i-1}, C = c) \neq 0$, then $\mathbb{P}(H_i = h_i | H_{i-1} = h_{i-1}, C=c)>0$ for all $h_i, h_{i-1} \in \mathscr{H}$ \citep{imai2004causal}.
    \item \cmtH{\emph{Weak ignorability}: Assume that for some certain $h \in \mathscr{H}$ and $c \in \mathscr{C}$, \\
    $\mathbb{E}_Y\left[ \lambda(x|H=h, C)\right] \indep H | C=c$. It follows that $\lambda_{h}(x) \indep H | C=c$ \citep{robins2000ignor,gelman2004applied,greenland2009identifiability}.}
    % Following \eqref{eq:ignor}, for some certain $h \in \mathscr{H}$ and $y \in \mathscr{Y}$, $\lambda_{h}(x) \indep H | C=c$ \citep{robins2000ignor,gelman2004applied,greenland2009identifiability}. }
    % \item \emph{Unconfoundedness}: 
    % For all $h \in \mathscr{H}$ and $c \in \mathscr{C}$, $\lambda_{h}(x) \indep H | C=c$.
    % Armed with this assumption, we can identify the causal effect within values of $C$ using adjustment formula \citep{neal2020introduction}, \ie,
    % \[
    %     \lambda_{h}(x) = \mathbb{E}_C\left[ \lambda(x|H=h, C)\right].
    % \]
\end{enumerate}
%\vspace{-0.05in}
Assumption 2 indicates that, for each event, each possible history has a non-zero probability of being assigned. Assumption 3, also called conditional exchangeability, means there are no unmeasured confounders. The user category is the only variable affecting both the treatment assignment and the next event that is present in the observational data set. Note that while assumption 3 is standard across all methods for estimating treatment effects, it is not testable in practice \citep{pearl2009causal,robins2000marginal}.

%\vspace{-0.1in}
\paragraph{Inverse Propensity Re-weighting}
Rather than directly analyzing the category-agnostic intensity, $\lambda_h$, our approach instead starts with establishing the \emph{category-agnostic probability density}, $f_h$, for the next event. 
This idea is summarized in the following Lemma~\ref{lemma:pseudo-prob}, which is based on the principle of inverse propensity treatment weight (IPTW) inspired by \cite{fitzmaurice2008longitudinal, robins1999association, wu2023counterfactual}:

\begin{lemma}[Category-agnostic probability density] % of MTPPs
    \label{lemma:pseudo-prob}
    \cmtH{Under the positivity and ignorability assumptions, the category-agnostic probability density function of the next event $x$ with its history $h$ is defined as} 
    \begin{equation}
        f_h(x) = \sum_{c \in \mathscr{C}} \frac{1}{\prod_{i=1}^{n+1} f(h_i|h_{i-1}, c)} f(x,h,c)
        % \int \frac{1}{\prod_{\tau=t-d+1}^{t} \prob{a_{\tau}|\ab_{\tau-1},\XB_{\tau}}} \prob{y,\ab,\XB} d\XB
        %= \int \frac{\mathbbm{1}\{\AB = \ab\}}{\prod_{\tau=t-d+1}^t \prob{A_\tau|\AB_{\tau-1}, \XB_{\tau}}} \prob{y, \AB, \XB}d \AB d \XB,
    \label{eq:counterfactual-pdf}
    \end{equation}
    where $h_0, h_1, \dots, h_n$ denote the previous user's history embedding trajectory and $p$ represents the probability density function of the data distribution. \cmtH{The joint distribution $f(x,h,c)$ is constructed from the DAG demonstrating the relationship between the parameters as shown in Figure~\ref{fig:causal-dag}.} The proof is included in Appendix~\ref{append:lemma-1}.

\end{lemma}
%\vspace{-0.05in}
Lemma~\ref{lemma:pseudo-prob} establishes a connection between what is observed and what could potentially occur, enabling us to express the category-agnostic probability density with re-weighted empirical samples. 
Thus we now propose a conditional intensity function $\lambda(\cdot|h)$ and use $\theta$ to denote all the parameters in the intensity $\lambda$ and the history encoder $h$. 
We approximate $\lambda_h$ using re-weighted data according to the lemma, leading to the following learning objective. The proof can be found in Appendix~\ref{append:prop-1}.

% \cmtH{move definition of $f_\theta$ here. connect to $\lambda_\theta$}

\begin{proposition}[Weighted maximum log-likelihood estimation]
    \label{prop:pesudo-objective}
    Let $\mathcal{D} = \{(x_i^{(k)}, \bar{h}_i^{(k)}, c)\}$ denote the set of observed data tuples, where $x_i^{(k)}$ is the $i$th observed event of the user $k$, $\bar{h}_i^{(k)} = (h_0^{(k)}, \dots, h_i^{(k)})$ is the corresponding history embedding trajectory, and $c$ is the user's category.
    For the sake of clarity, we use the notation of $h(x)$ in place of $h$ to represent the up-to-date history embedding, acknowledging that both terms have been used interchangeably in earlier discussions.
    The learning objective can be approximated by maximizing the following weighted log-likelihood:
    % \begin{equation}
    \begin{align}
    \begin{split}
        % \ell_w(\mathcal{D};\theta) =~
        & \mathbb{E}_{h} ~\left[\mathbb{E}_{x \sim p_{h}} \left[\log f(x|h(x))\right]\right] \\
        \propto~&\frac{1}{K}\sum_{(x_i^{(k)},\bar{h}_i^{(k)},c) \in \mathcal{D}} w(\bar{h}_i^{(k)}, c) \log f(x_i^{(k)}|h_i^{(k)})\\
        % =\text{not correct}~& \frac{1}{K}\sum_{k=1}^K \left( \int_{\mathscr X} \log \bigg (w(\bar{h}, c) \lambda_\theta^{(k)}(x|h) \bigg) d \mathbb N^{(k)}(x) -  \int_{\mathscr X}  \lambda_\theta^{(k)}(x|h) dx \right).\\
        % &= \frac{1}{K}\sum_{k=1}^K w(\bar{h}_i^{(k)}, c) \left( \int_{\mathscr X} \log \bigg ( \lambda_\theta^{(k)}(x|h) \bigg) d \mathbb N^{(k)}(x) -  \int_{\mathscr X}  \lambda_\theta^{(k)}(x|h) dx \right)\\
        =~& \frac{1}{K}\sum_{k=1}^K \Big( \int_{\mathscr X} w(\bar{h}, c) \log \big ( \lambda(x|h(x)) \big) d \mathbb N^{(k)}(x) 
         -\int_{\mathscr X} w(\bar{h}_n^{(k)}, c) \lambda(x|h^{(k)}(x)) dx \Big), 
    \end{split}
    \label{eq:weighted-log-likelihood}
    \end{align}
    where $K$ is the number of users, $\mathbb{N}^{(k)}$ is the counting measure for each user indexed by $k$, respectively. Furthermore, $\bar{h}_n^{(k)}$ denotes the set of observed history embedding for the $k$th user before $t(x)$.
    % \cmtH{Both $\bar{h}_n^{(k)}$ and $h^{(k)}$ seem to have to be defined here as this is where multiple users are defined. $f_\theta$ has been moved to Section 2.}
    % \woody{$f_\theta$ is the corresponding conditional probability function of $\lambda_\theta$ according to \eqref{eq:cond-prob}.
    % In the second integral, $\bar{h}_n$ denotes the observed history embeddings before $h$.
    % TODO: Might as well need to explain $h^{(k)}$ here...}
    % \woody{We might need to define some of the concepts prior to the proposition. Otherwise, the proposition is too long. }
    Here $w$ denotes the subject-specific IPTW, which takes the form:
    \begin{equation}
        w(\bar{h},c) = \frac{1}{\prod_{i=1}^{n+1} f(h_i|h_{i-1}, c)}.
        \label{eq:iptw}
    \end{equation}
    % It is obtained by converting the causal DAG into the joint probability function following the work of \cite{robins2000marginal}.  
\end{proposition}

%\vspace{-0.15in}
\paragraph{Conditional History Transition}

To learn the model by maximizing \eqref{eq:weighted-log-likelihood}, one needs to specify the conditional transition probability $f(h_i|h_{i-1}, c)$ between two consecutive history embedding variables $h_i$ and $h_{i-1}$ given the user's category $c$. Following \cite{naimi2014constructing}, we propose a non-parametric binning method to obtain the probability $f(h_i|h_{i-1}, c)$. 
It comes with three practical benefits: ($i$) Discretized sample space enables the probability to be discontinuous and identifiable, assisting the positivity assumption; ($ii$) Bin size can be adjusted to help stabilize weights; ($iii$) Binning method provides a less complex yet still effective way to estimate probabilities.

% The proposed method comes with a few benefits in practice: ($i$) The conditional history transition model discretizes the sample space and allows the probability estimated to be discontinuous so it becomes identifiable. It makes the second assumption easier to reach in practice; ($ii$) The bin size can be tuned which helps stabilize the magnitude of weights; ($iii$) The binning method estimates the probability by counts instead of complex modeling. Performance is improved even if binning is coarse.

% \begin{enumerate}
%     \item The conditional history transition model discretizes the sample space and allows the probability estimated to be discontinuous so it becomes identifiable. This provides flexibility to model $f(h_i|h_{i-1}, c)$ as not all transition combinations for $h_i$ and $h_{i-1}$ will take place or be observed. This makes the second assumption on positivity easier to reach in practice.
%     \item The size of the bins can be adjusted manually given the number of samples and how spread out they are. This helps ensure there are few bins where a small number of samples lead to very large scores, further stabilizing the magnitude of weights.
%     \item The binning method estimates the probability by counting the number of samples in each bin and does not require complex model fitting. With the inverse propensity weights added to the log-likelihood function, we expect improvement in the prediction performance even if the binning is coarse and only produces rough approximates. 
%     % It is therefore computationally efficient.
% \end{enumerate}

Our method is implemented as follows. We partition the unit history embedding space $\mathscr{H}$ uniformly into $1/\delta$ bins of equal size of $\delta$, denoted by $\mathscr{H}_u \in \mathscr{H}$. 
Without loss of generality, we can treat the category space as a continuous space as well. 
We can therefore also partition the cluster space $\mathscr{C}$ into $R$ bins of equal size, denoted by $\mathscr{C}_r$. 
In this work, however, the cluster space is already discrete. Then the observed conditional history transition space $\Omega \coloneqq \mathscr{H} \times \mathscr{H} \times \mathscr{C}$ can be constructed as a set of $(1/\delta)^2\times R$ elements, where each of them can be denoted by $\mathscr{B}(u, u', r) \coloneqq \mathscr{H}_u \times \mathscr{H}_{u'} \times \mathscr{C}_r \subseteq \Omega$. Here, $u$ and $u'$ are indices of the set $\mathscr{H}$ and $r$ is the index of the set $\mathscr{C}$. The count of observed conditional history transition tuples in an arbitrary bin $\mathscr{B}$ can be expressed as $\sum_{k=1}^K \sum_{j=1}^{N_T^{(k)}} \mathbf{1}[(h_j^{(k)}, h_{j-1}^{(k)}, c^{(k)}) \in \mathscr{B}(u, u', r)]$,
where $\mathbf{1}$ denotes the indicator function. 

\begin{figure}[!t]
\centering
\begin{subfigure}[b]{.32\linewidth}
\includegraphics[width=\linewidth]{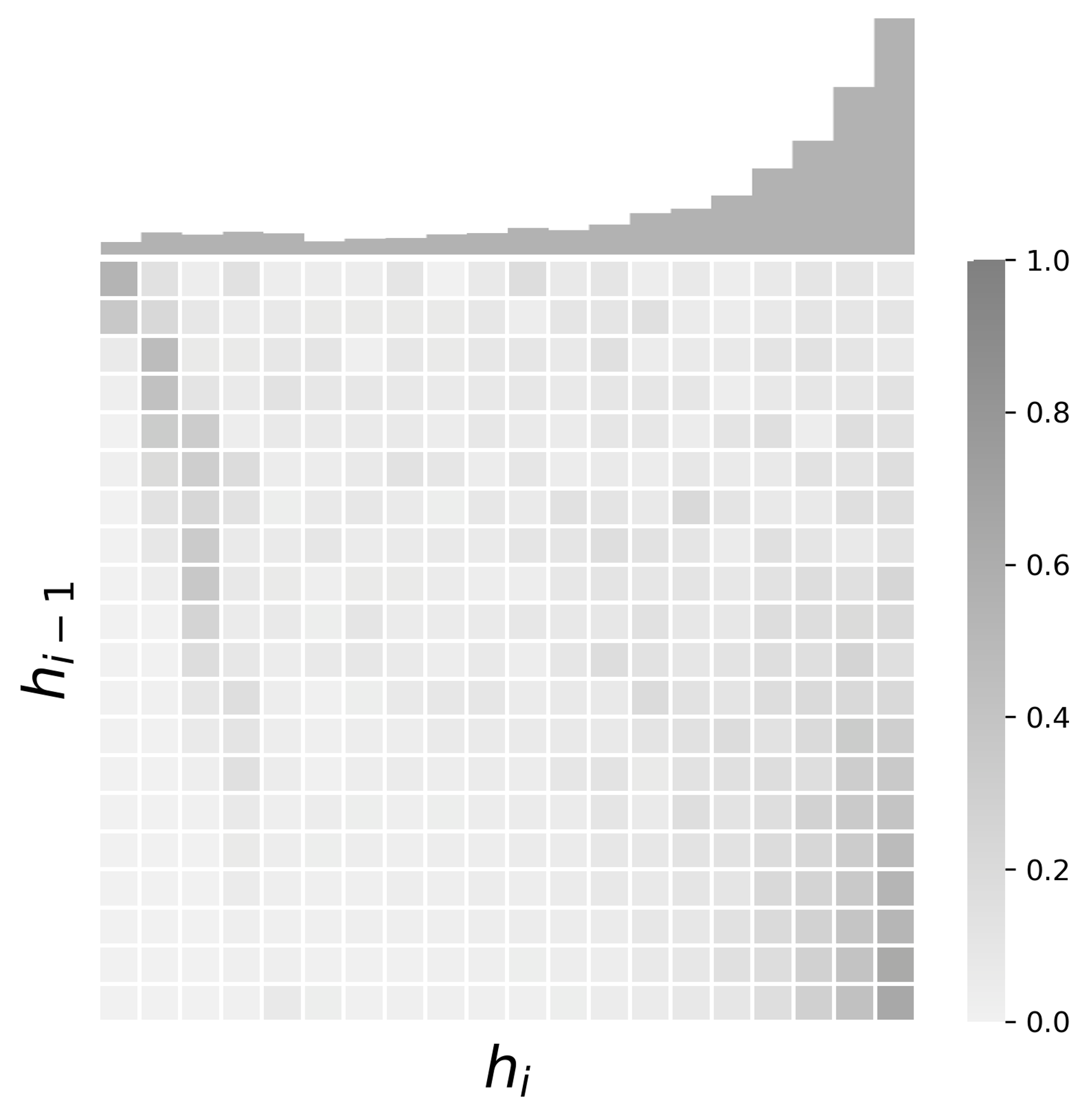}
\caption{$C=1$}
\end{subfigure}
\hfill
\begin{subfigure}[b]{.32\linewidth}
\includegraphics[width=\linewidth]{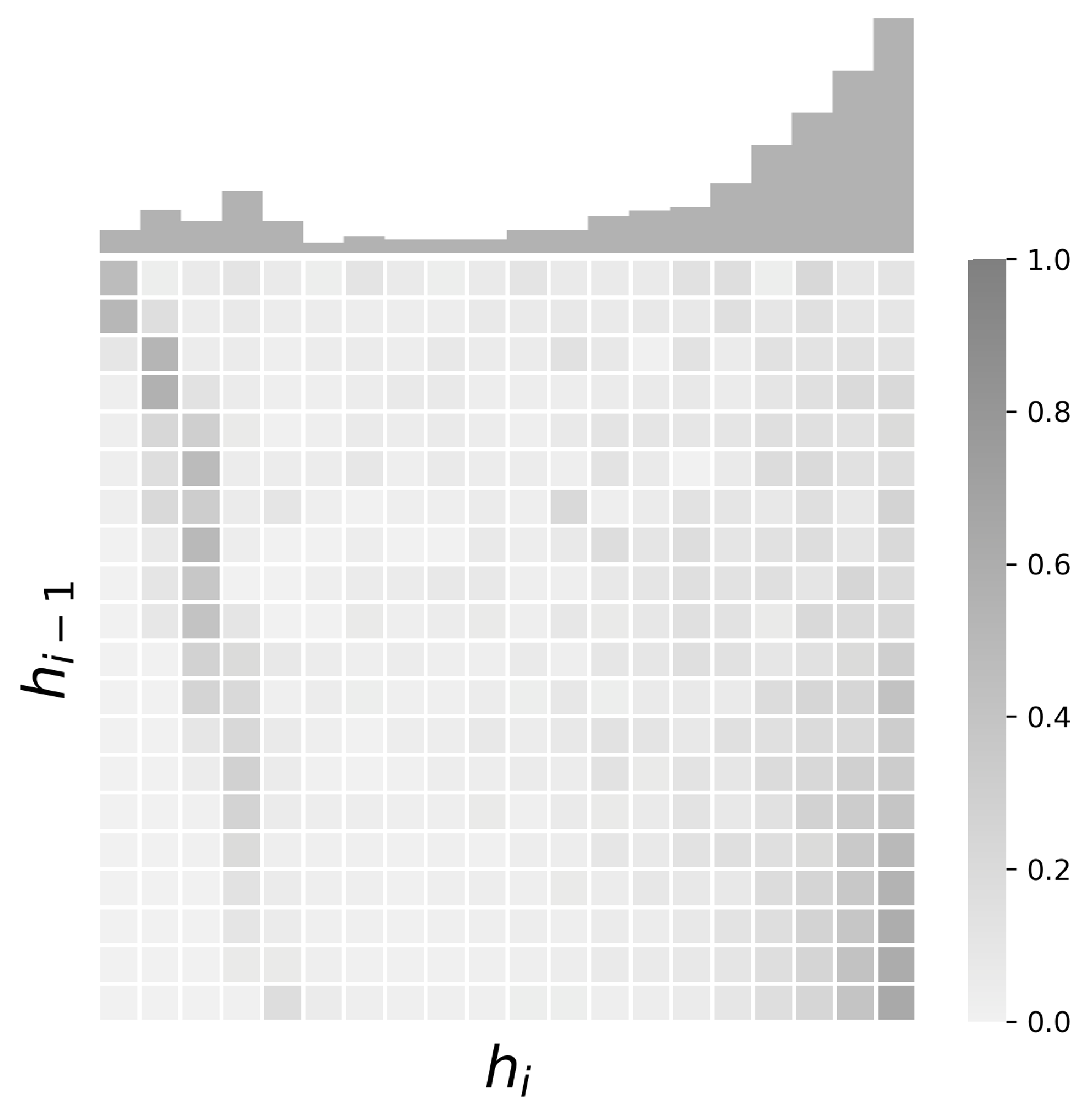}
\caption{$C=2$}
\end{subfigure}
\hfill
\begin{subfigure}[b]{.32\linewidth}
\includegraphics[width=\linewidth]{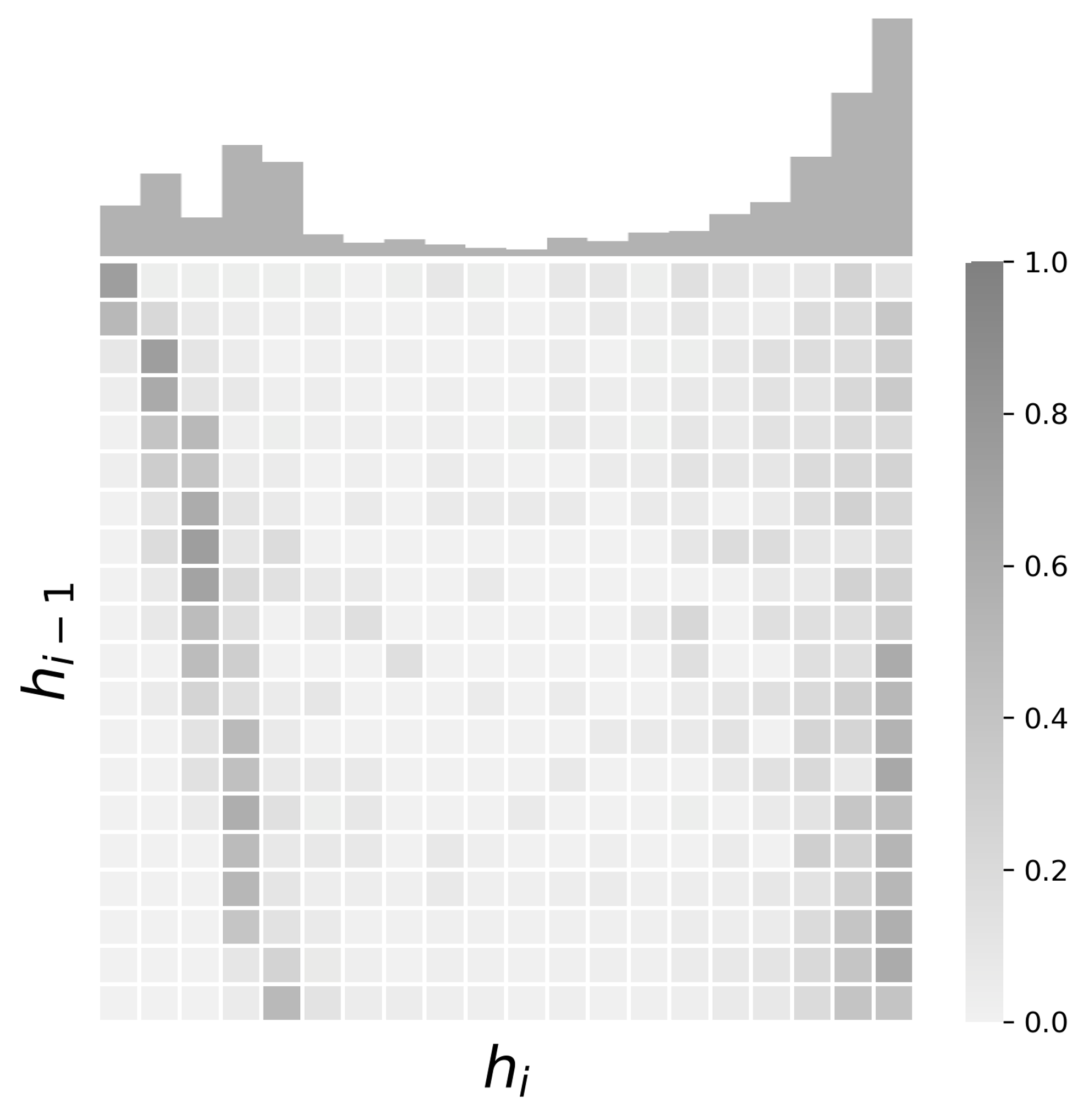}
\caption{$C=3$}
\end{subfigure}
\caption{
% The histogram of $p(h_i | C)$ and the conditional transition probability $p(h_i | h_{i-1}, C)$ in three categories $C=1, 2, 3$ when $1/\delta=20$. 
The histogram of $p(h_i | C)$ and the conditional transition probability $p(h_i | h_{i-1}, C)$ for the three categories $C=1, 2, 3$ when $1/\delta=20$ visually reveal distinct patterns. 
% We can visually identify the difference of the conditional histogram and transition probability under different user category $C$, indicating their distinctive pattern of user behavior.
These differences in the conditional transition probabilities across user categories highlight their unique behavior pattern.
}
\label{fig:cond-trans-prob}
%\vspace{-0.1in}
\end{figure}

We can thus obtain a discretized histogram across all bins $\mathscr{B}(u, u', r)$. 
The probability of an arbitrary tuple $(h, h', c')$ falling in the bin $\mathscr{B}(u, u', r)$ can be approximated by the proportion of the counts:
\begin{align}
\begin{split}
\widehat{p}(u, u', r) =
\frac{\sum_{k=1}^K \sum_{j=1}^{N_T^{(k)}} \mathbf{1}[(h_j^{(k)}, h_{j-1}^{(k)}, c^{(k)}) \in \mathscr{B}(u, u', r)]}
{\sum_{k=1}^K N_T^{(k)}},
\label{eq:bin_prob}
\end{split}
\end{align}
\cmtH{where $u$ and $u'$ refer to the bin indices for the transition in $h$, and $r$ is the index for category.} An example of the conditional transition probabilities and the histograms is visualized in Fig.~\ref{fig:cond-trans-prob}.
% Then the joint probability $f(h_i,h_{i-1},c)$ can be 
% which can be denoted by $p_\delta(h_i,h_{i-1},c)$ since it depends on the bin size. It is an approximation of the true joint probability:
% \begin{equation}
% f(h_i,h_{i-1},c) \approx \sum_l p_l \cdot \mathbbm{1}[(h_i, h_{i-1}, c) \in \mathscr{B}_{l}].
% \label{eq:bin_prob_approx}
% \end{equation}
Therefore, the conditional transition probability can be estimated using \eqref{eq:bin_prob} by 
% when $(h_i, h_{i-1}, c) \in \mathscr{B}_{l}$ can be approximated by the probability of samples falling in a specific bin
\begin{align}
\begin{split}
\widehat{f}_\delta(h_i|h_{i-1},c) =
\frac{\sum_{(u, u', r)} \widehat{p}(u, u', r) \cdot \mathbf{1}[(h_i, h_{i-1}, c) \in \mathscr{B}(u, u', r)]}{ \sum_{u=1}^{1/\delta} \widehat{p}(u, u', r) \cdot \mathbf{1}[h_i \in \mathscr{H}_u]}. 
\label{eq:prob_approx}
\end{split}
\end{align}
With the transition probability estimated, we can approximate the IPTW weights following \eqref{eq:iptw}:
\begin{equation}
    \widehat{w}_\delta(\bar{h}, c) = \frac{1}{\prod_{i=1}^{n+1} \widehat{f}_\delta(h_i|h_{i-1}, c)}.
\label{eq:w_delta}
\end{equation}
% The method offers an efficient non-parametric way to obtain the weights, and does not require the transition probability to be continuous. It makes the approximations more flexible to model the causality which itself is discontinuous as well, making the second assumption for counterfactual intensity on positivity more practical. Additionally, by offering a way to adjust the discretization of the sample domain, it allows the weighted to be further stabilized to an appropriate degree. 
% The binning method offers. When the number of bins gets too small, then the fidelity of the approximate degrades as well. On the other hand, too many bins will provide a more accurate approximation, but it will result in significant computational power by the scale of $q^{2p}$.
This method is limited by the dimensionality of the hidden history embedding denoted by $q$, since the number of bins grows quickly with it. The complexity of obtaining each conditional transition probability scales with $q$, and that of bin construction scales exponentially.
% \woody{We might need an illustrative example here to explain how we get the equation 10. }

%\vspace{-0.1in}
\paragraph{Sensitivity Analysis}
% \label{sec:sensitivity}

We now look into how the discretization via binning affects the log-likelihood estimation and then study quantitatively its effects on the history transition probabilities. We denote the true IPTW values calculated based on $f(h_i|h_{i-1},c)$ as $w$ and define $\ell^*(w)$ as the maximal log-likelihood with optimal model parameters $\theta^*$ with the weights $w$. With discretization over the history embedding space, we denote the weights derived from $\widehat{f}_\delta(h_i|h_{i-1},c)$ by $\widehat{w}_\delta$. Then $\ell^*(\widehat{w}_\delta)$ refers to the maximal log-likelihood given $\theta^*$ when the weights take the value of $\widehat{w}_\delta$.

\begin{proposition}[Improvement in binning]
Let the history embedding space $\mathscr{H}$ be uniformly partitioned into bins of equal size $\delta$, we denote the corresponding IPTW derived by $\widehat{w}_\delta$. We have
\begin{equation*}
     \ell^*(\widehat{w}_1) \le \ell^*(\widehat{w}_\delta) \le \ell^*(w), \quad \forall \delta \in (0,1]. 
\end{equation*} 
\label{prop:improve}
%\vspace{-0.25in}
\end{proposition}
The binning approximation will inevitably introduce bias into the log-likelihood $\ell(\widehat{w}_\delta)$ since the weights are approximated in \eqref{eq:w_delta}. The binning procedure approximates the joint probability $f(h_i,h_{i-1},c)$ by $\widehat{f}_\delta(h_i,h_{i-1},c)$ as in \eqref{eq:prob_approx}. We analyze the bias and variance introduced by binning in Appendix~\ref{append:sensitivity}. We further assume that when the number of events gets sufficiently large in a sequence, the relationship between events stabilizes and the joint probability function $f(h_i,h_{i-1},c)$ becomes symmetrical between $h_i$ and $h_{i-1}$. This leads to $\partial f/\partial h_i=\partial f/\partial h_{i-1}=f'$.
%Here we assume that when the number of events gets sufficiently large in a sequence, the relationship between events stabilizes and the joint probability function $f(h_i,h_{i-1},c)$ becomes symmetrical between $h_i$ and $h_{i-1}$. This enables the simplification of the partial derivatives which we denote by $f'$:
% \[
% \frac{\partial f}{\partial h_i}=\frac{\partial f}{\partial h_{i-1}} = f'.
% \]
\begin{proposition}[Optimal bin size]
Following the
%the same assumptions by Lemma \ref{append:lemma-bias} and the aforementioned 
symmetry assumption of partial derivatives, the optimal bin size can be obtained by minimizing the \cmtH{integrated minimum square error (IMSE)} between the true probability density function $f$ and the estimator by binning $\hat{f}$ given $m$ samples of data tuples. The proof is in Appendix~\ref{append:sensitivity}. %When $\int f'\neq 0$, the optimal bin size is as follows:
\begin{align*}
\begin{split}
\delta^* = \left( \frac{2}{\left( \int f^{'} \right)^2}  \right)^{1/4} m^{-1/4} &\text{ if } \int f'\neq 0;\\
\delta^* = \left( \frac{4}{\int f^{'^2}}  \right)^{1/5} m^{-1/5} &\text{ if }\int f' =0.
\end{split}
\end{align*}
% When $\int f' =0$, the optimal bin size is as follows:
% \[
% \delta^* = \left( \frac{4}{\int f^{'^2}}  \right)^{1/5} m^{-1/5}.
% \]
\label{append:opt-delta}
\end{proposition}
%\vspace{-0.2in}
% \cmtH{Another approach to this can be found in \cite{hahn2022feature}.}

%\vspace{-0.05in}
\subsection{Learning and Inference}
%\vspace{-0.05in}

To learn the point process model, we fit the model parameters $\theta$. This is conducted by optimizing the weighted log-likelihood function in \eqref{eq:weighted-log-likelihood}, where the IPTW values $w(\bar{h},c)$ need to be obtained separately. The weights are instead derived from conditional history transition in \eqref{eq:iptw}. It indicates there is cyclic dependence between the weights $w(\bar{h},c)$ and the model parameters. Therefore, to obtain estimates for both $\theta$ and $w(\bar{h},c)$, it is necessary to conduct an alternate optimization. 

Given the objective log-likelihood function \eqref{eq:weighted-log-likelihood} and the IPTW expression \eqref{eq:w_delta}, we propose a joint learning algorithm to obtain estimates for both the model parameters $\hat{\theta}$ and the IPTW values $\widehat{w}_\delta$ in Appendix~\ref{sec:algorithm} with details on implementation. 
The prediction and inference of the proposed model follow the standard point processes inference method \citep{zhu2021spatio}, which is further detailed in Appendix~\ref{sec:inference}.

\begin{table*}[!t]
\centering
\caption{Prediction MAE of synthetic experiments. 
% \woody{We need the results of two other key baselines. Here \texttt{R-} represents randomly choosing a model for the use behavior prediction.}
}
\resizebox{.95\linewidth}{!}{%
\begin{tabular}{cccc:ccc:cc}
\hline
\hline
                     % & \multicolumn{6}{c}{MAE per method}                                                   \\
Syn Exp No. & \texttt{R-NH} & \texttt{NH} & \textcolor{red}{\texttt{C-NH}} & \texttt{R-RMPTT} & \texttt{RMTPP}  & \textcolor{red}{\texttt{C-RMTPP}} & \texttt{Exp Hawkes} & \texttt{Self-Corr} \\ \hline
1 & 1.6545
% TPP fixed distribution per cluster           
& 1.0308 & 0.9145 &2.5738& 1.9304 & 1.8751         &  2.8332    & 2.6224                 \\
2 & 1.5544
% TPP fixed distribution per cluster-A1         
& 1.0019 & 0.9611 &2.2110& 1.7429 & 1.7010         &  1.5854    & 2.3097                 \\
3 & 1.6760
% TPP fixed distribution per cluster-A2
& 0.9217 & 0.9055 &2.2965& 1.8711 & 1.8128         &  1.2508    & 1.9449                 \\
4 & 1.4612
% TPP varying distribution per cluster
& 1.2222 & 0.9986 &2.0820& 1.5833 & 1.5542         &  1.4748    & 1.8293      
    \\
5 & 2.0417
% TPP different distribution family per cluster 
& 1.4299 & 1.4067 &2.4207& 2.0234 & 1.9868         &  2.1177    & 2.7442                 \\
6 & 2.5801
% MTPP fixed distribution per cluster           
& 1.0796 & 1.0523 &2.3872& 1.9197 & 1.8127         &  n/a    & n/a                 \\ \hline\hline
\end{tabular}
}
\label{tbl:results_sim}

\end{table*}

\begin{figure}[!t]
\centering
\begin{subfigure}[b]{.32\linewidth}
\includegraphics[width=\linewidth]{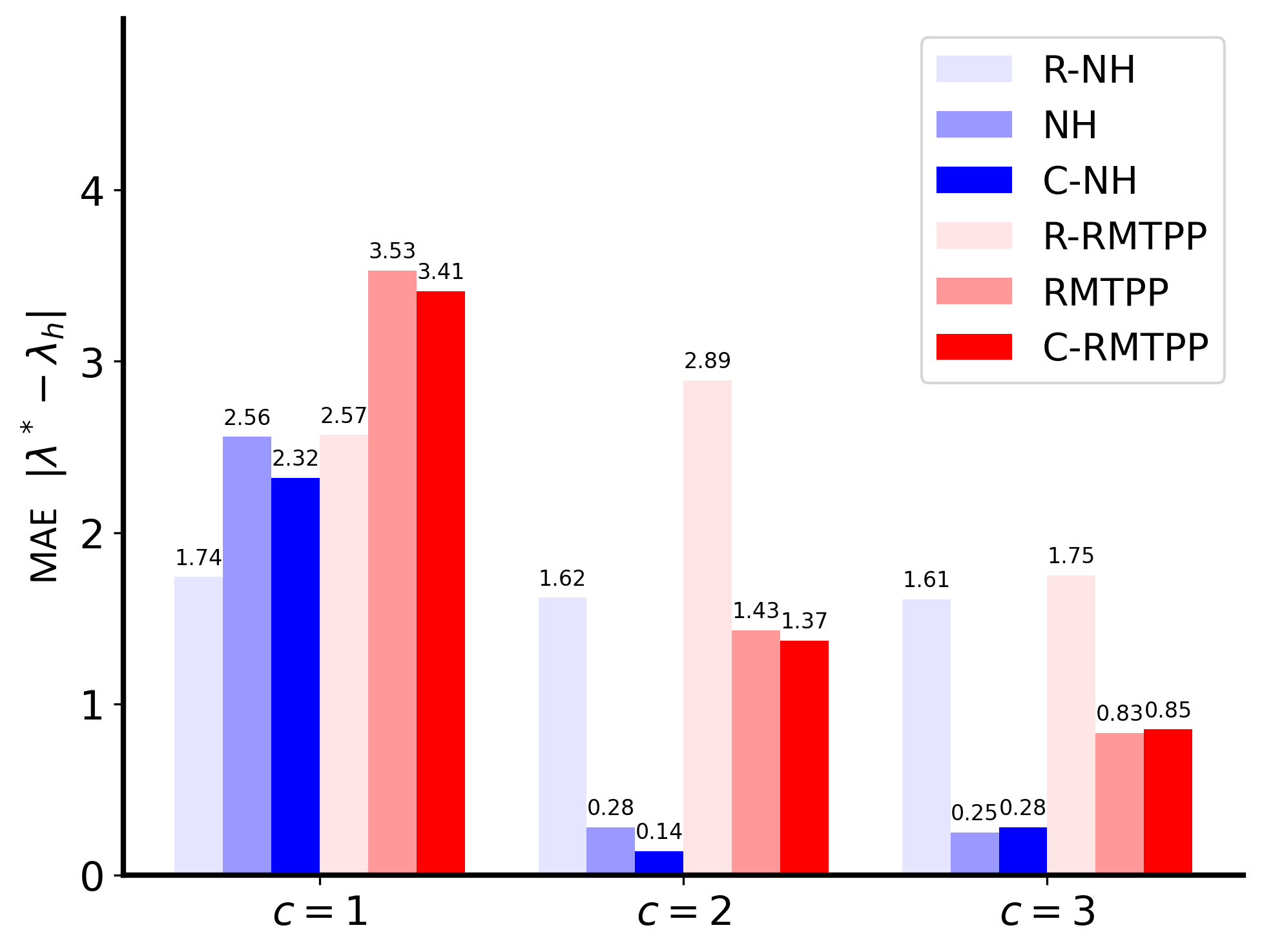}
\caption{Syn Exp No.1}
\end{subfigure}
\begin{subfigure}[b]{.32\linewidth}
\includegraphics[width=\linewidth]{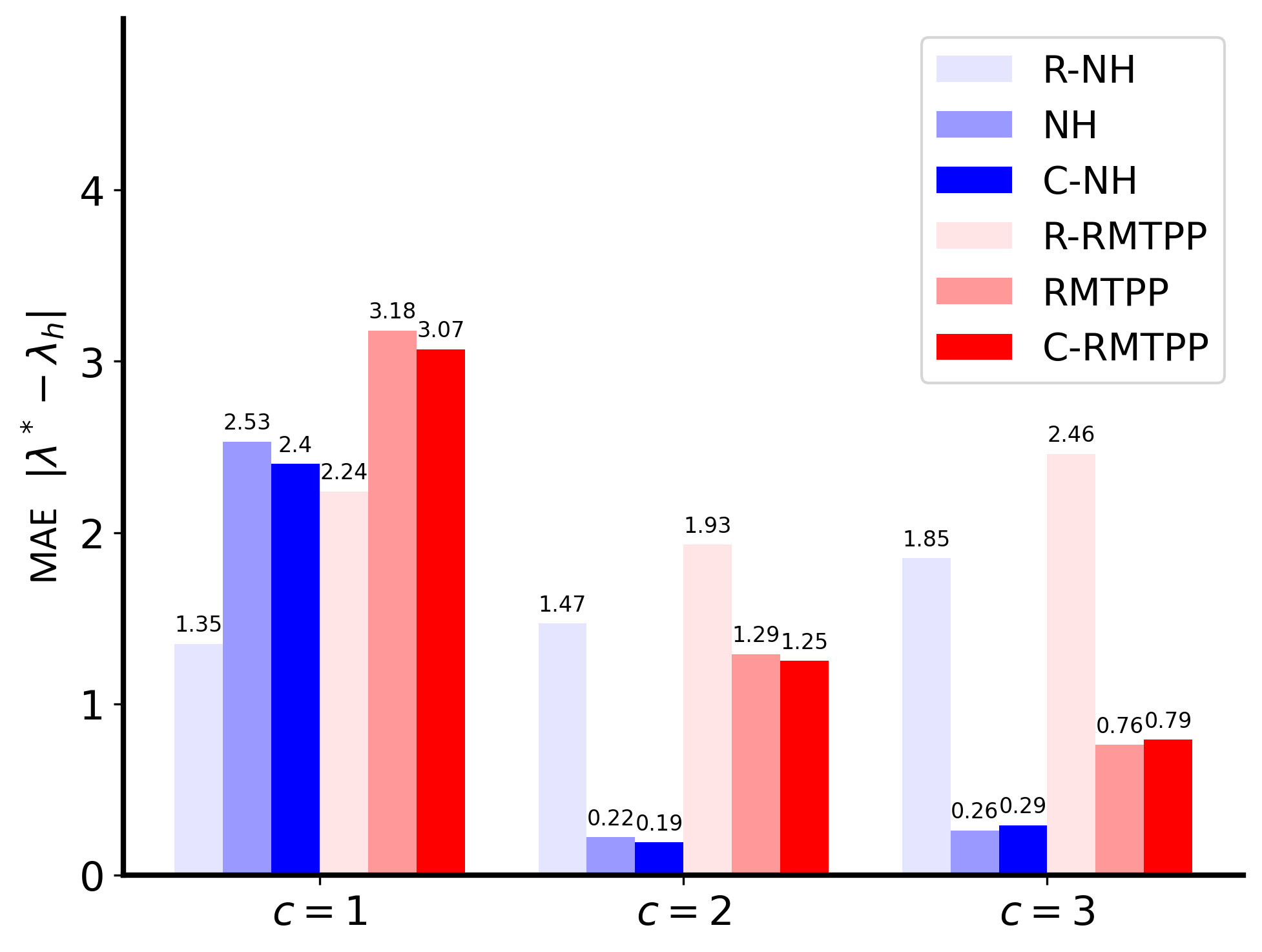}
\caption{Syn Exp No.2}
\end{subfigure}
% \vfill
\begin{subfigure}[b]{.32\linewidth}
\includegraphics[width=\linewidth]{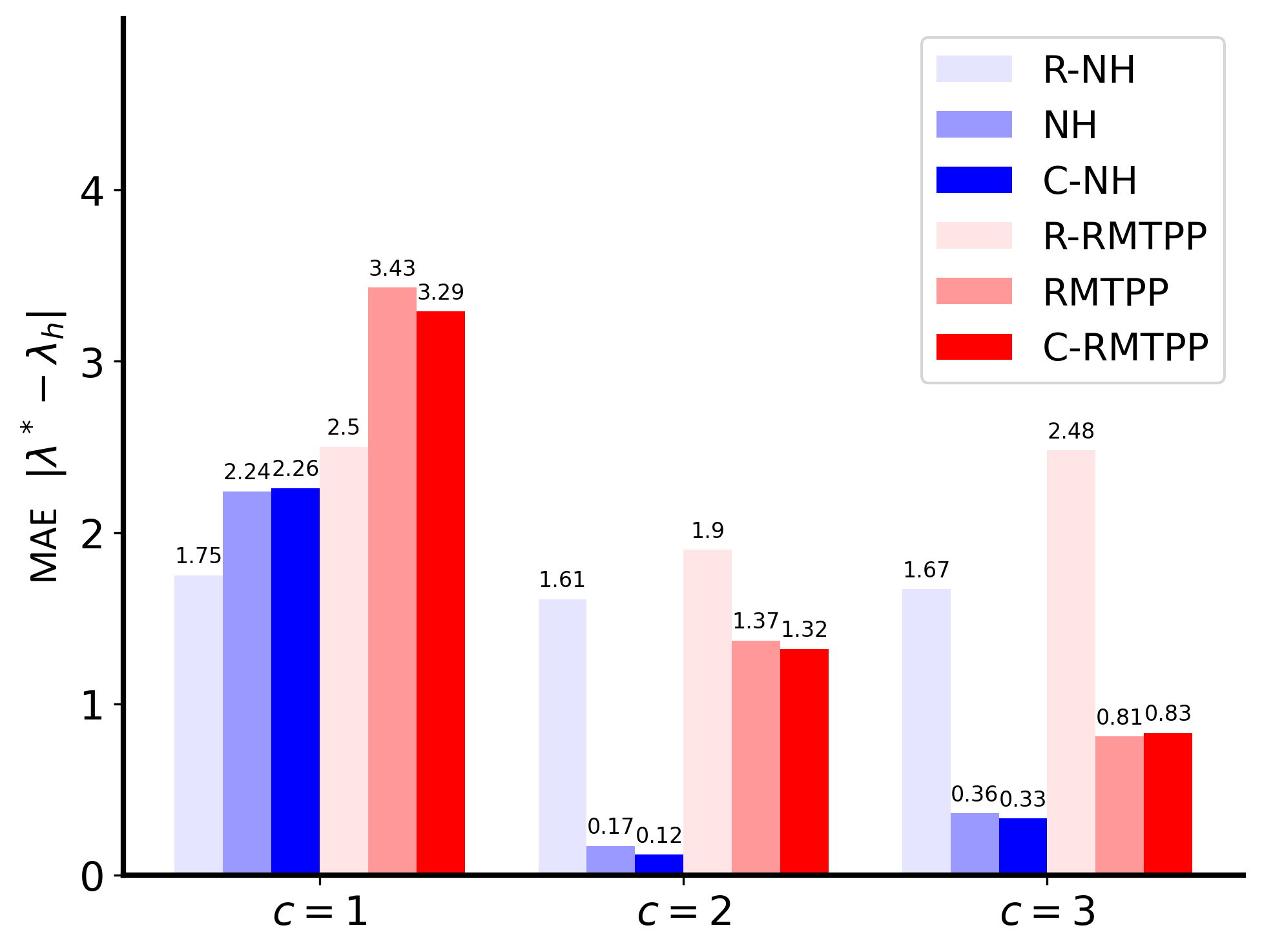}
\caption{Syn Exp No.3}
\end{subfigure}
\vfill
\begin{subfigure}[b]{.32\linewidth}
\includegraphics[width=\linewidth]{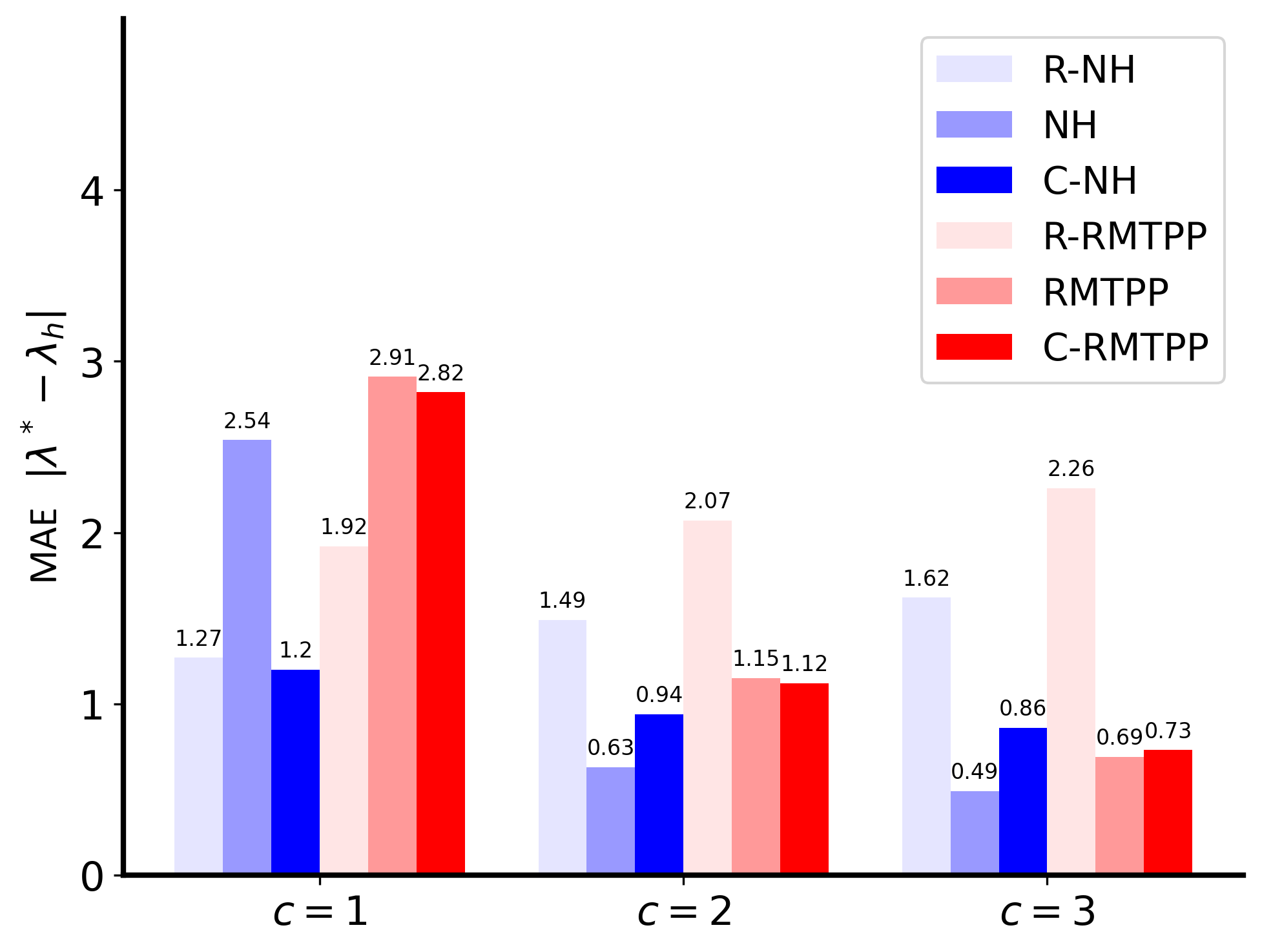}
\caption{Syn Exp No.4}
\end{subfigure}
% \vfill
\begin{subfigure}[b]{.32\linewidth}
\includegraphics[width=\linewidth]{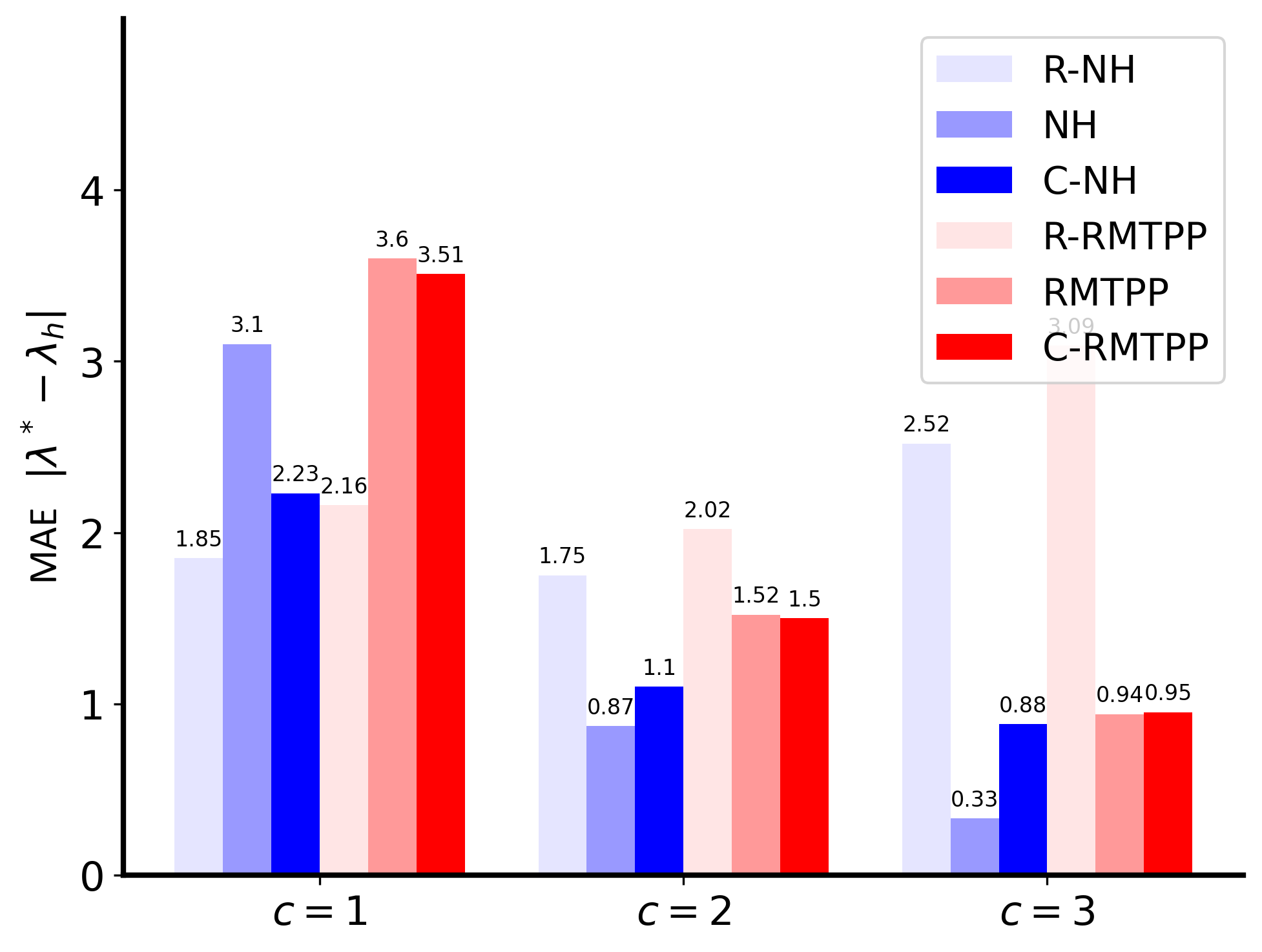}
\caption{Syn Exp No.5}
\end{subfigure}
\begin{subfigure}[b]{.32\linewidth}
\includegraphics[width=\linewidth]{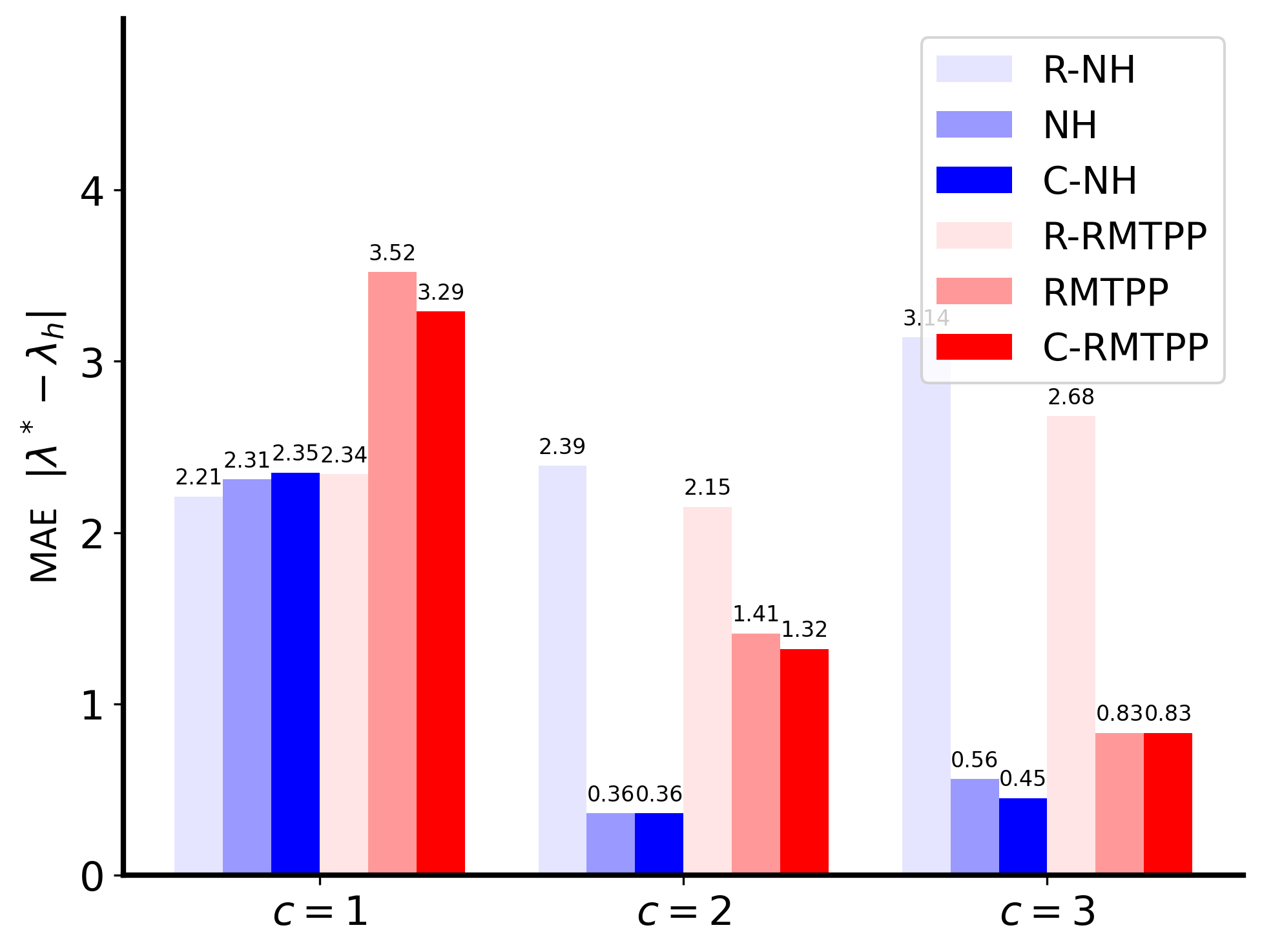}
\caption{Syn Exp No.6}
\end{subfigure}
\caption{Category-wise prediction MAE of synthetic experiments.}
\label{fig:MAE_barchart}
%\vspace{-0.1in}
\end{figure}

%\vspace{-0.15in}
\section{Experiments}
\label{sec:experiments}
%\vspace{-0.1in}
To evaluate the efficacy of the proposed counterfactual event prediction framework, we apply it in a set of simulation studies, a pair of hyper-parameter studies, and two real-world applications. We compare the performance of the proposed framework in these studies against several popular point processes as baseline models. All simulations are conducted on Google Colab free edition with 12.7 GB of RAM, each taking between 15-75 minutes to finish.
%\vspace{-0.1in}
\subsection{Synthetic Studies}
\label{sec:sim}
%\vspace{-0.05in}
% We first study the performance of the framework in a range of realistic scenarios, then we carry out an ablation study on two key model hyper-parameters.
% In the simulation studies, we first investigate the performance of the proposed framework in a range of scenarios mimicking what typically takes place in reality. This part of the study ensures the proposed method is capable of accommodating common situations. We also carry out an ablation study of the framework looking into how the model's hyper-parameters can impact the performance in event prediction.
\paragraph{Experimental Setup} 
% \cmtH{1. Experiment setup, what data generation scheme etc. 2. Baseline methods (what baseline is, why it is included, what is the difference,) and evaluation metric. (MAE on lambda, MAE on prediction for real data, Acc on real data.)}
% \woody{Discuss evaluation metrics and baselines methods in this paragraph.}
% In this simulation study, we generate user event sequence data from simple temporal point process models with variables for the conditional intensity function specified. To account for the multiple categories, we utilize different point process models to generate sequences from each category. These sequences are then divided into training and testing data respectively. Here we specify that there are three categories and generate 1,200 event sequences for training, and a further 300 for testing from all three of them. 
Throughout the simulation study, we obtain training data and testing data by generating them from known point process models. Both the training and the testing data sequences are generated from $|\mathscr{C}|=3$ categories each corresponding to a different point process model or a set of similar models. The events are confined in the time range of $[0, T), T=100$. %For example, sequences from each cluster can be sampled from a distinct point process model such that the user behavior between clusters is different. 
For testing, the category for the sequences is taken out. We obtain 1,200 sequences for the training set and an additional 300 for the testing set. The IPTW values are truncated at $10^6$ to prevent instability.
% \woody{mentioning the weight truncation in the setup...}
%We conduct multiple experiments to account for various scenarios corresponding to potential use cases in the real world. 
% Across the experiments, we evaluate the proposed framework against a set of baseline methods on its prediction performance. 
%Additionally, we carry out an ablation study to investigate the effects of parameters of our model on its performance.

% For evaluation, we compare the proposed framework against the baselines using the same generated datasets. Specifically, we look into the following baselines: Neural Hawkes point process \cite{mei2017neural}, RMTPP \cite{du2016recurrent}, standard exponential Hawkes point process, and self-correcting point process. Our proposed methods include the Counterfactual-Neural Hawkes (C-NH) and Counterfactual-RMTPP (C-RMTPP), which build upon the two baselines utilizing the counterfactual weighting scheme. 

\emph{Baselines}: We elect to utilize four point process models commonly utilized in event sequence data modeling, including Neural Hawkes (\texttt{NH}) method, Recurrent Marked Temporal Point Process (\texttt{RMTPP}) model, temporal Hawkes process with exponential triggering function (\texttt{Exp Hawkes}), and self-correcting point process (\texttt{Self-corr}). Additionally, we utilize Neural Hawkes and RMTPP for the case when users in the testing set are \textit{randomly} assigned their category, providing two more baselines \texttt{R-NH} and \texttt{R-RMTPP}.
% The baseline methods cover a wide range of different approaches to point process modeling from the basic Hawkes model to the state-of-the-art ones utilizing hidden embedding variables including \texttt{NH} and \texttt{RMTPP}. 
For our counterfactual framework, we construct Counterfactual-Neural Hawkes (\texttt{C-NH}) and Counterfactual-RMTPP (\texttt{C-RMTPP}) such that we can directly evaluate the performance between \texttt{NH} and \texttt{C-NH} and for \texttt{RMTPP} as well. 
% Additionally, the self-correcting point process is included given its different triggering structure and serves as an additional reference in baseline. 

\emph{Evaluation metrics}: We elect to utilize the mean absolute error (MAE) of the predicted conditional intensity as the error metric in the testing set. Since in this study the conditional intensity functions for the sequence-generating models are specified and known, we directly compare the true conditional intensity against the estimate with parameters fitted by each method. For each event sequence in the testing set, we are able to calculate the error between the two functions across the full range of time, constructing the MAE value. 

\emph{Synthetic data}: We investigate six synthetic experiments. The first is a base experiment with training and testing sequences evenly split between three categories each generated by an exponential Hawkes point process with fixed model parameters. The second and the third experiments change the composition of the training data. The fourth looks into exponential Hawkes PP with the model parameter randomly drawn. In the fifth experiment, sequences for the third category are generated by a Neural Hawkes PP instead. In the final case, the temporal PP models generating sequences are extended to contain a discrete mark variable. Detailed descriptions are in Appendix~\ref{sec:simulation-detail}.

\paragraph{Results}
% \cmtH{Discuss Table 1, Fig. 4}
Results are shown in Table~\ref{tbl:results_sim}. The performance by $\texttt{C-NH}$ and $\texttt{C-RMTPP}$ are generally better than counterparts by 2-10\% and 2-5\% respectively, with the former usually having smaller errors due to more complex model structure. But there are cases where the counterfactual methods underperform on MAE for individual categories as shown in Fig~\ref{fig:MAE_barchart}. This is due to the proposed method balancing between categories instead of preferring one of them. There is one exception for the fourth experiment, where the MAE of $\texttt{C-NH}$ is 20\% smaller. This can be attributed to that the event pattern of each category is no longer fixed. \texttt{R-NH} and \texttt{R-RMTPP} tend to have higher errors but the difference between categories is much smaller since they are randomly assigned. The other two baselines tend to have higher MAE values, but \texttt{Exp Hawkes} sometimes have smaller errors than \texttt{C-RMTPP}. This is partly due to that the event sequences in the study are generated from exponential Hawkes PP themselves. Overall the performance of the proposed counterfactual methods is consistently better.

\begin{table}[!t]
\centering
    \caption{MAE {\it w.r.t.} the number of bins $1/\delta$.}
    \resizebox{.5\linewidth}{!}{
        \begin{threeparttable}
        \begin{tabular}{ccc:cc}
        \hline\hline
        \# bins ($1/\delta$) & \textcolor{red}{\texttt{C-NH}} & Pct.     & \textcolor{red}{\texttt{C-RMTPP}} & Pct.     \\ \hline
        1\tnote{*}             & 1.0155                         & 0\%     & 1.6559                            & 0\%     \\
        5             & 0.9347                         & -8.0\%  & 1.5442                            & -6.7\%  \\
        10            & 0.9145                         & -10.0\% & 1.5029                            & -9.2\%  \\
        15            & 0.9066                         & -10.7\% & 1.4788                            & -10.6\% \\
        20            & 0.9012                         & -11.3\% & 1.4691                            & -11.3\% \\ \hline\hline
        \end{tabular}
        \begin{tablenotes}
        \item[*]{\footnotesize $1/\delta=1$ leads to no discretization and all weights taking the value of 1, reverting the method to its unweighted version.
        }
        \end{tablenotes}
        \end{threeparttable}}
        \label{tbl:ablation_U}
\end{table}

\begin{figure}[!t]
%\begin{wrapfigure}{R}{1\linewidth}
% \vspace{-0.15in}
\centering
\begin{subfigure}{0.49\linewidth}
\includegraphics[width=\linewidth]{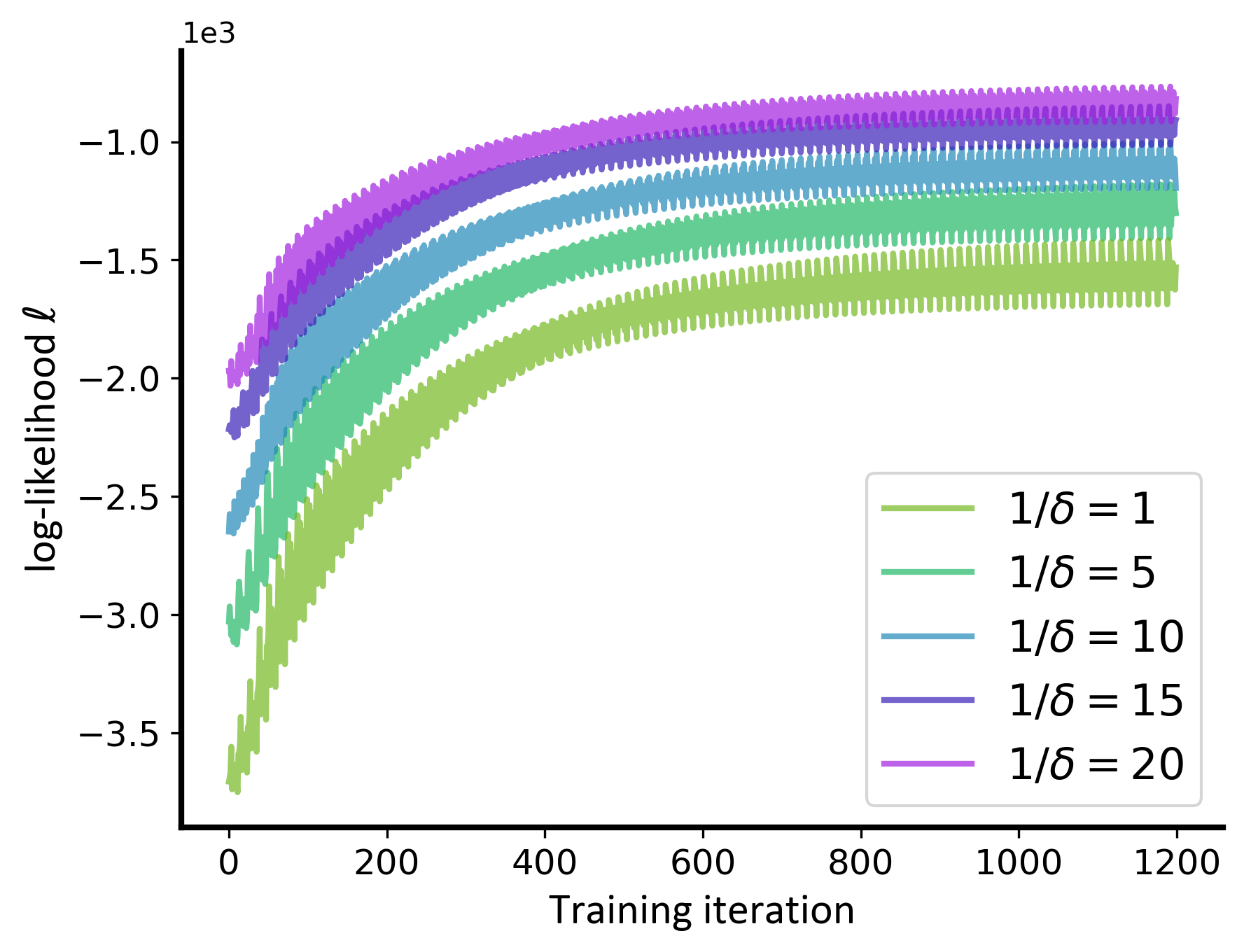}
%\vspace{-0.2in}
\caption{The number of bins $1/\delta$.}
\end{subfigure}
\begin{subfigure}{0.49\linewidth}
\includegraphics[width=\linewidth]{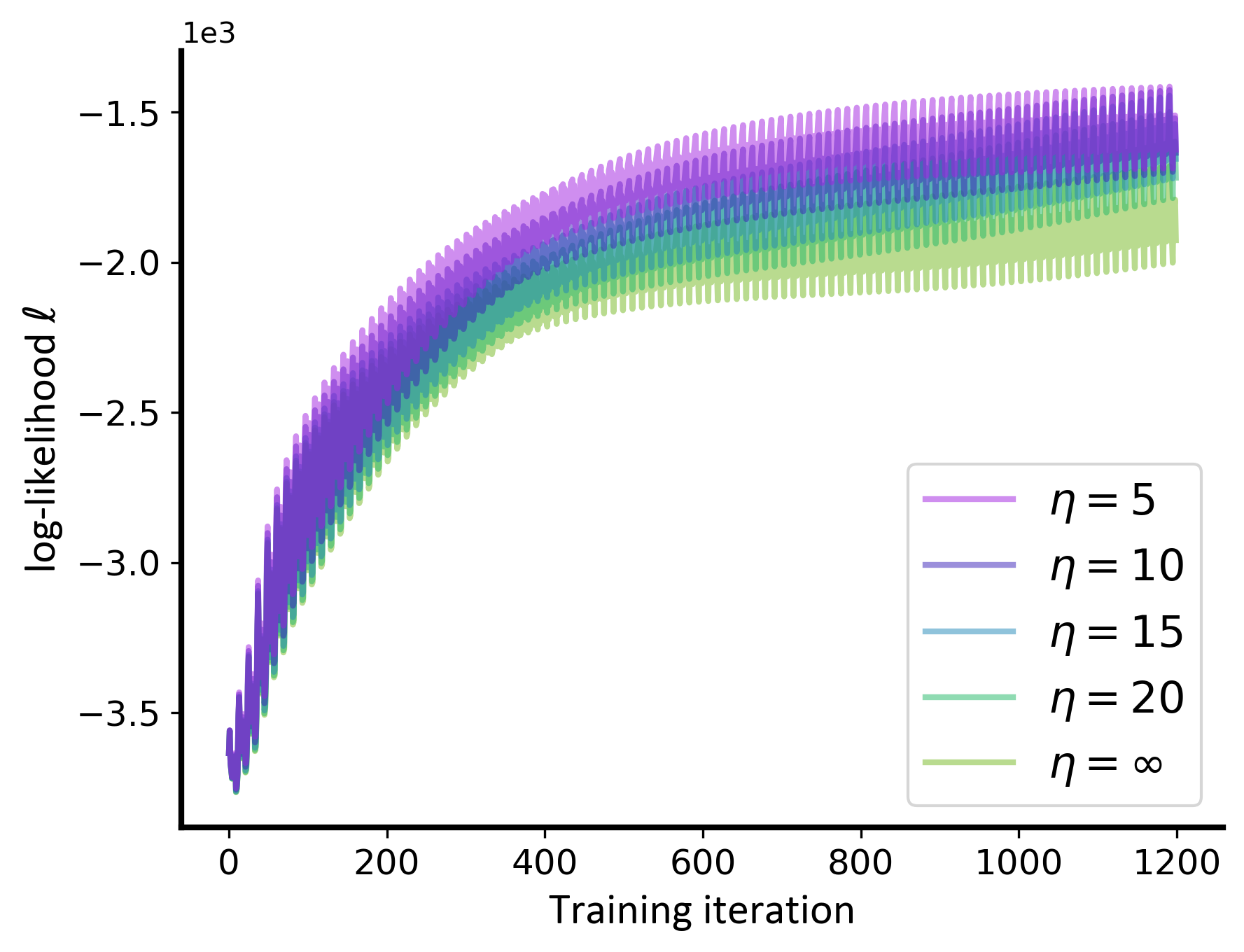}
%\vspace{-0.2in}
\caption{The number of epochs $\eta$.}
\end{subfigure}
%\vspace{-0.2in}
\caption{Convergence analysis.}
\label{fig:ablation_plot}
%\vspace{-0.2in}
%\end{wrapfigure}
\end{figure}

%\vspace{-0.1in}
\paragraph{Hyper-parameter Selection}
% \label{sec:ablation}
% \cmtH{Ablation study: what we do, why we do that, what is our purpose, what metric we look for}
% \cmtH{discuss through fig.5,6, Table 2,3}

% \woody{discuss how to select U and E in the end.}
% Beyond the synthetic studies, we now look more closely into the hyper-parameters of the counterfactual framework and how they affect the performance of the method in prediction. 
The proposed framework utilizes two hyper-parameters that can be tuned: ($i$) the number of bins used to approximate the conditional transition probability, denoted by $1/\delta$, and ($ii$) the number of epochs at which pace the IPTW weights are updated, denoted by $\eta$.
% In short, $\eta$ inversely reflects the frequency of re-fitting the conditional transition probabilities. In this ablation study, 
We investigate how they influence the prediction performance using the same data setting from Synthetic Exp. 1.
% these two model parameters influence the prediction performance of the model. We generate 1,200 training sequences and 300 testing sequences from three Hawkes process models with fixed parameters evenly such that each category contributes 400 training and 100 testing sequences, the same setting as Synthetic Exp. 1. 
We compare the two counterfactual methods \texttt{C-NH} and \texttt{C-RMTPP} with their counterparts \texttt{NH} and \texttt{RMTPP}. We fix the dimension of embedding $q=1$, and evaluate them using the error metric MAE. Additionally, we plot the change of training log-likelihood $\ell$ over 100 training epochs to study the convergence behavior.

% \woody{Shorten the following two paragraphs...}
%\vspace{-0.05in}
\emph{Number of bins $1/\delta$}: 
We alter the number of bins $1/\delta$ used to approximate conditional history transition probabilities while fixing $\eta=5$.
%We study the effect on the proposed method when we alter the number of bins used to approximate conditional history transition probabilities. We fix the number of epochs $\eta=5$ and vary $\delta$ to take several different values. We also include the extreme case when $\delta=1$, which corresponds to no discretization takes place and all weights remain at 1, \ie, becoming the unweighted baselines. 
Results are shown in Table~\ref{tbl:ablation_U} and the convergence behavior is plotted in Fig.~\ref{fig:ablation_plot} (a). 
We observe that as $1/\delta$ increases the MAE for both \texttt{C-NH} and \texttt{C-RMTPP} decrease consistently. However, it is slow and relatively insignificant. 
We can deduce that while finer discretization improves prediction performance, there is a trade-off between the improvement and the additional computational cost when $1/\delta$ gets larger.

\begin{table}[!t]
\centering
    \caption{MAE {\it w.r.t.} the number of epochs $\eta$.}
    \resizebox{.5\linewidth}{!}{
        \begin{threeparttable}
        \begin{tabular}{ccc:cc}
        \hline\hline
        \# epochs ($\eta$) & \textcolor{red}{\texttt{C-NH}} & Pct.   & \textcolor{red}{\texttt{C-RMTPP}} & Pct.   \\ \hline
        5               & 0.9388                         & -7.5\% & 1.5570                            & -6.0\% \\
        10              & 0.9455                         & -6.9\% & 1.5630                            & -5.6\% \\
        15              & 0.9560                         & -5.9\% & 1.5833                            & -4.4\% \\
        20              & 0.9683                         & -4.6\% & 1.5995                            & 3.4\%  \\
        $\infty $\tnote{*}        & 1.0155                         & 0\%    & 1.6559                            & 0\%    \\ \hline\hline
        \end{tabular}
        \begin{tablenotes}
        \item[*]{\footnotesize $\eta=\infty$ leads to no update of IPTW with all staying at 1, reverting the method to its unweighted version.
        }
        \end{tablenotes}
        \end{threeparttable}}
        \label{tbl:ablation_E}
        %\vspace{-0.1in}
\end{table}

\emph{Number of epochs $\eta$}:
We alter the number of epochs $\eta$ at which pace the IPTW weights are fitted while fixing $1/\delta=5$.
% We look into the effect on the model when we alter the number of epochs $\eta$ at which pace the IPTW weights are fitted. We fix the number of bins to $1/\delta=10$ and vary $\eta$ to construct several test cases. We include the extreme case when $\eta=\infty$, which corresponds to when the IPTW weights are never updated after being initialized to 1. This is effectively the unweighted case. 
Results are shown in Table~\ref{tbl:ablation_E} and the convergence behavior is plotted in Fig.~\ref{fig:ablation_plot} (b). 
We observe as $\eta$ gets smaller the MAE for both methods decreases. 
% This is reasonable since a lower frequency of updating the IPTW weights is expected to lead to worse prediction performance. 
However, the relative change in MAE is smaller compared to that in $1/\delta$. 
We deduce it is unnecessary to have a very small $\eta$ for the model to remain effective, and some computation can be saved by tuning it adequately.

\begin{table*}[!t]
\centering
\caption{Numerical Results for real case studies.}
\resizebox{0.9\linewidth}{!}{
\begin{threeparttable}
\begin{tabular}{ccccc:ccccc}
\hline\hline
                                       % &                                     & \multicolumn{4}{c}{Method}                     \\
Real Exp                                & Metric    & \texttt{R-NH}  & \texttt{NH}     & \textcolor{red}{\texttt{C-NH}} & \texttt{R-RMTPP}& \texttt{RMTPP}  & \textcolor{red}{\texttt{C-RMTPP}}  \\ \hline
\multirow{2}{*}{Netflix}        & MAE &6.2633& 5.6598 & 5.6323 &6.3853& 5.6529 & 5.6278     \\
                                       & Acc.\tnote{1} &0.160&0.293  & 0.307 &0.147& 0.267  & 0.347         \\ \hline
\multirow{2}{*}{Amazon} & MAE&14.3511& 4.3830 & 4.2291 &4.2643& 4.2526 & 4.1690     \\
                                       & Acc.\tnote{1} &0.172&0.238  & 0.274 &0.138& 0.190  & 0.226     \\ \hline\hline
\end{tabular}
\begin{tablenotes}
\item[1]{\footnotesize Acc. refers to the accuracy predicting the actual discrete intent on whether it is included in the five most likely intents predicted by each method.
}
\end{tablenotes}
\end{threeparttable}
}
\label{tbl:real-case}

\end{table*}

\subsection{Real Case Studies}
\label{sec:applications}
%\vspace{-0.05in}
% \woody{Opening. Use the same baselines from the synthetic data. Metrics. Transition to data and its results.} 

We evaluate two real-world applications: Netflix rating prediction and Amazon seller contact prediction. They deal with event sequence data with a discrete mark variable describing the type of events. We use the same methods as before. Details of data and settings are in Appendix~\ref{sec:real-detail}.

% We utilize the same four baseline methods as the synthetic experiments, and the two proposed counterfactual methods \texttt{C-NH} and \texttt{C-RMTPP}. For details of dataset and settings, refer to Appendix \ref{sec:real-detail}.

\emph{Evaluation metrics}: Since the true conditional intensity is no longer available, we propose new error metrics.
% we now utilize different error metrics for performance evaluation and comparison.
For time prediction, we calculate the error between the predicted time and the actual time of occurrence for the next event, then calculate the mean absolute error (MAE). 
For mark prediction, we calculate the accuracy of the top five predicted marks against the actual mark for the next event.

%\vspace{-0.1in}
\paragraph{Netflix Ratings} The Netflix ratings data is a public dataset that records the user ratings on movies \citep{bennett2007netflix}. 
Users rate movies from different genres, and we treat the rating records for each user as an event sequence. The training set contains 300 sequences including movies from 45 genres, assigned to ten categories. An extra 75 sequences without category information are obtained as the testing set.
% Typically users rate multiple movies they watch from different genres over a long period of time, often a few years. We regard the rating records from each user as an event sequence. Thus we obtain from the dataset a series of user rating sequences containing the time and genre of the corresponding movies. The training set contains 300 user sequences containing movies from 45 genres. The users are put into ten different categories. A further 75 user sequences are obtained as the testing set without category information. 
The results are shown in Table~\ref{tbl:real-case}.
% \emph{Results}: 
% The 100 users are divided into 6 different clusters. For prediction, we further obtain the event history of 25 additional users. We assume the cluster/group information of these 25 users are unknown, given the lack of clear inclinations towards a certain movie genre. Similar to the simulation studies, we carry out the MAE between the predicted and the actual time of the immediate next rating event. We utilize the NH model with our proposed method for comparison. For the mark prediction, which is the combination of movie genre and sentiment of the rating, we compare the proportion of correct predicted marks between the two method.
The results indicate the performance of our proposed methods is better than the baselines both in terms of time prediction and mark prediction. For time prediction, the proposed methods achieve a 0.5\% improvement over their counterparts, partly due to the smaller size of the dataset. For mark accuracy, the improvement is larger at 5\% and 30\% for \texttt{C-NH} and \texttt{C-RMTPP} respectively. The random baselines are worse than the others.

% \begin{table}[!t]
% \centering
% \caption{Numerical Results for real case studies.}
% \resizebox{1\linewidth}{!}{
% \begin{threeparttable}
% \begin{tabular}{ccccc:ccccc}
% \hline\hline
%                                        % &                                     & \multicolumn{4}{c}{Method}                     \\
% Real Exp                                & Metric    & \texttt{R-NH}  & \texttt{NH}     & \textcolor{red}{\texttt{C-NH}} & \texttt{R-RMTPP}& \texttt{RMTPP}  & \textcolor{red}{\texttt{C-RMTPP}}  \\ \hline
% \multirow{2}{*}{Netflix}        & MAE &6.2633& 5.6598 & 5.6323 &6.3853& 5.6529 & 5.6278     \\
%                                        & Acc.\tnote{1} &0.160&0.293  & 0.307 &0.147& 0.267  & 0.347         \\ \hline
% \multirow{2}{*}{Amazon} & MAE&14.3511& 4.3830 & 4.2291 &4.2643& 4.2526 & 4.1690     \\
%                                        & Acc.\tnote{1} &0.172&0.238  & 0.274 &0.138& 0.190  & 0.226     \\ \hline\hline
% \end{tabular}
% \begin{tablenotes}
% \item[1]{\footnotesize Acc. refers to the accuracy predicting the actual discrete intent on whether it is included in the five most likely intents predicted by each method.
% }
% \end{tablenotes}
% \end{threeparttable}
% }
% \label{tbl:real-case}
% %\vspace{-0.2in}
% \end{table}

%\vspace{-0.1in}
\paragraph{Amazon Seller Contact} The Amazon seller contact data \citep{dong2023ascisdata} records how Amazon selling partners utilize the support contact system. Each event contains the time and the intent out of 117 possible choices. 
The training set contains 3,000 sequences assigned to 12 categories, with a further 500 sequences without category obtained for testing.
% Our second application looks into how Amazon conducts customer support for sellers. We obtain from the dataset \cite{dong2023ascisdata} 3,000 user event sequences each from a selling partner. Each event contains the time of the event and a discrete mark for the intent out of 117 possible choices. The 3,000 users for training fall into ten categories, and we further obtain 500 user event sequences without category information for testing.
% \emph{Results}: 
From the results, we observe our proposed methods perform better as well. For time prediction, the MAE is lower by 4\% and 2\% respectively for \texttt{C-NH} and \texttt{C-RMTPP}. The mark accuracy is higher by 15\% and 19\% as well.

\section{Conclusions}
%\vspace{-0.1in}
% \vspace{-0.2in}
In this work, we have investigated the user-event prediction problem, where we aim to conduct event prediction for new users of limited history without category information.
%to predict the upcoming events for users given past event sequences history from them and others.
We addressed the challenge of unbiased prediction for ``new'' users through the proposed counterfactual framework, which reweights data using IPTW.
It is evaluated using a series of simulation studies and two real-world cases, demonstrating improvement in prediction performance.
%We validated the efficacy of our proposed framework using a series of simulation studies, which shows the reweighting scheme improves the prediction performance of point process models consistently utilizing hidden history variables.
% This work has shown that by introducing the weights from conditional transition probabilities, bias can be reduced in the prominent point process models that encode historical information using a set of hidden variables. 
There are two limitations of this work. We have only applied the framework to lower-dimensional hidden variables due to computational constraints, but they can be high-dimensional as well. Additionally, we have assumed that category information is always given in the training set, which may not be the case in reality. We plan to further investigate efficient categorization (clustering) based on event sequences under such circumstances.

\bibliographystyle{plain}
\bibliography{references}

\newpage
\appendix
\section{Derivation of Point Process Conditional Probability}
\label{append:derivation-cond-prob}
% \vspace{-0.05in}
% Assume that we have total number of $\mathbb{N}([0, T] \times \mathcal{M})$ observations in $\boldsymbol{x}$. For any given $t \in [0, T]$, we assume that $n$ events happened before $t$ and denote the occurrence time of the latest event as $t_n$. 
% Let $\Omega = [t, t + dt) \times B(s, ds)$ where $s \in \mathcal{S}$. 
The conditional probability of point processes can be derived from the conditional intensity \eqref{eq:cond-intensity}. Suppose we are interested in the conditional probability of events at a given point $x \in \mathcal{X}$, and we assume that there are $i$ events that happen before $t(x)$. Let $\Omega(x)$ be a small neighborhood containing $x$. According to \eqref{eq:cond-intensity}, we can rewrite $\lambda(x|\mathcal{H}_{t(x)})$ as following:
\[
    \begin{aligned}
        \lambda(x|\mathcal{H}_{t(x)}) &= \mathbb{E}\left( d\mathbb{N}(x) | \mathcal{H}_{t(x)} \right) / dx = \mathbb{P}\{x_{i+1} \in \Omega(x) | \mathcal{H}_{t(x)}\} /dx\\
        &= \mathbb{P}\{x_{i+1} \in \Omega(x)| \mathcal{H}_{t_{i+1}} \cup \{t_{i+1} \geq t(x)\}\} /dx\\
        &= \frac{\mathbb{P}\{x_{i+1} \in \Omega(x), t_{i+1} \geq t(x) | \mathcal{H}_{t_{i+1}}\}/dx}{\mathbb{P}\{t_{i+1} \geq t(x) | \mathcal{H}_{t_{i+1}}\}}. \\
        % &= \frac{f(t, s)}{1 - F(t)}
    \end{aligned}
\]
Here $\mathcal{H}_{t_{i+1}} = \{x_1, \dots, x_i\}$ represents the history up to $i$-th events.
If we let $F(t(x)|\mathcal{H}_{t(x)}) = \mathbb{P}(t_{i+1} < t(x) | \mathcal{H}_{t_{i+1}})$ be the conditional cumulative probability, and $f(x|\mathcal{H}_{t(x)}) \triangleq f(x_{i+1} \in \Omega(x)|\mathcal{H}_{t_{i+1}})$ be the conditional probability density of the next event happening in $\Omega(x)$. Then the conditional intensity can be equivalently expressed as
\[
    \lambda(x|\mathcal{H}_{t(x)}) = \frac{f(x|\mathcal{H}_{t(x)})}{1 - F(t(x)|\mathcal{H}_{t(x)})}.
\]
We multiply the differential $dx = dtdm$ on both sides of the equation and integral over the mark space $\mathcal{M}$:
\[
    \begin{aligned}
        dt \cdot \int_{\mathcal{M}}\lambda(x|\mathcal{H}_{t(x)})dm &= \frac{dt \cdot \int_{\mathcal{M}}f(x|\mathcal{H}_{t(x)})dm}{1 - F(t(x)|\mathcal{H}_{t(x)})} = \frac{dF(t(x)|\mathcal{H}_{t(x)})}{1 - F(t(x)|\mathcal{H}_{t(x)})} \\ &= -d\log{(1 - F(t(x)|\mathcal{H}_{t(x)}))}.
    \end{aligned}
\]
Hence, integrating over $t$ on $[t_i, t(x))$ leads to the fact that
\[
\begin{aligned}
    F(t(x)|\mathcal{H}_{t(x)}) &= 1 - \exp \left (-\int_{t_{i}}^{t(x)}\int_{\mathcal{M}} \lambda(x|\mathcal{H}_{t(x)})dmdt \right ) \\ 
    &= 1 - \exp \left (-\int_{[t_i, t(x)) \times \mathcal{M}} \lambda(x|\mathcal{H}_{t(x)})dx \right)
\end{aligned}
\]
because $F(t_{i}) = 0$. Then we have
\[
    f(x|\mathcal{H}_{t(x)}) = \lambda(x|\mathcal{H}_{t(x)}) \cdot \exp \left (-\int_{[t_i, t(x)) \times \mathcal{M}} \lambda(x|\mathcal{H}_{t(x)})dx \right ),
\]
which corresponds to \eqref{eq:cond-prob}.
% The joint p.d.f. for a realization is then, by the chain rule, $f(x_1, ..., x_{\mathbb{N}([0, T]\times \mathcal{S})}) = \prod_{i=1}^{\mathbb{N}([0, T]\times \mathcal{S})} f(t_i, s_i)$. Then the log-likelihood of an observed sequence $\boldsymbol{x}$ can be written as
% \[
%     l(\boldsymbol{x}) = \sum_{i=1}^{\mathbb{N}([0, T]\times \mathcal{S})} \log \lambda(t_i, s_i) - \int_{0}^{T}\int_{\mathcal{S}} \lambda(\tau, u)dud\tau .
% \]

The log-likelihood of one observed event series in \eqref{eq:log-likelihood} is derived, by the chain rule, as
\[
    \begin{aligned}
        \ell(x_1, \dots, x_{N_T}) &= \log f(x_1, \dots, x_{N_T}) = \log \prod_{i=1}^{N_T}f(x_i|\mathcal{H}_{t_i}) \\
        &= \int_{\mathcal{X}} \log f(x|\mathcal{H}_{t(x)})d\mathbb{N}(x) \\
        &= \int_{\mathcal X}  \log \lambda(x| \mathcal{H}_{t(x)} ) d \mathbb N(x) -  \int_{\mathcal X}  \lambda(x| \mathcal{H}_{t(x)} ) dx.
    \end{aligned}
\]
The log-likelihood of $K$ observed event sequences can be conveniently obtained with the counting measure $\mathbb{N}$ replaced by the counting measure $\mathbb{N}_k$ for the $k$-th sequence.

% in (\ref{eq:sequence-likelihood})
\section{Markov History Process}
\label{sec:markov}
The history embedding used by NPP defined in \eqref{eq:history-var} can be regarded as a series of realizations of $q$-dimensional homogeneous Markov process \citep{kuo2018markov}.
% \woody{We start our discussion by showing }
Notably, the history $\mathcal{H}_{t(x)}$ in \eqref{eq:nn-lam} is a \emph{filtration} associated with the MTPP, which is a sequence of $\sigma$-algebras (or information sets) that satisfy $\mathcal{H}_{s} \subseteq \mathcal{H}_{t}$ for $s \leq t$, representing the accumulation of information up to the time $t$.
Statistically, the history embedding $h(x)$ can be viewed as the realization of a $q$-dimensional random variable that effectively encapsulates the information contained in the filtration.
To facilitate the explanation for the rest of the paper, the history embedding up to the $i$th event, or $h(x_i)$; and the one for the next unobserved event $x$, or $h(x)$, is denoted as $h_i$ and $h$, respectively.
It is important to note that while $h(x)$ and $h$ can be continuously defined in $\mathscr{X}$, $h_i$ is only constructed at the observed events. 
This work investigates the relation between the history embedding $h_i$ and $h$, and the corresponding intensity carefully to construct the counterfactual intensity framework. 

\section{Proof of Lemma~\ref{lemma:pseudo-prob}}
\label{append:lemma-1}
The counterfactual probability density of MTPPs can be derived from the joint probability of discrete event $x$, history embedding $h_i$, and the category $c$. From the causal graph shown in Fig.~\ref{fig:causal-dag}, we can formulate the joint probability function $f(x,h,c)$ as follows:
\begin{equation*}
f(x,h,c)  = f(x|h,c) \cdot f(c) \cdot \prod_{\tau=1}^i f(h_\tau|h_{\tau-1}, c) \cdot f(h_0|c).
\label{eq:joint-pdf}
\end{equation*}
From the joint probability distribution, we can then derive the conditional distribution $f(x|h)$, or shorthanded as $f_{h}(x)$:
\begin{equation*}
\begin{split}
f_{h}(x) & = \int f(x|h, c) \cdot f(c) dc\\
& = \int f(x|h, c) \cdot \frac{\prod_{\tau=1}^i f(h_\tau|h_{\tau-1}, c)}{\prod_{\tau=1}^i f(h_\tau|h_{\tau-1}, c)} \cdot f(c) dc\\
& = \int \frac{1}{\prod_{\tau=1}^i f(h_\tau|h_{\tau-1}, c) f(h_0|c)} \left[ f(x|h, c) \cdot \prod_{\tau=1}^i f(h_\tau|h_{\tau-1}, c) \cdot f(c) \cdot f(h_0|c)  \right] dc.
\end{split}
\end{equation*}
After reorganizing the conditional distribution $f_{h}(x)$, we plug in the expression for the joint distribution:
\begin{equation*}
f_{h}(x) = \int \frac{1}{\prod_{\tau=1}^i f(h_\tau|h_{\tau-1}, c)  f(h_0|c)} f(x,h,c) dc.
\end{equation*}
In the setting of our problem, the initial history embedding $h_0$ starts where there have yet been any events observed at all. Therefore, $f(h_0|c)$ can be regarded as a constant of $1$ since $h_0$ is independent from $c$ and not random. This results in the following expression:
\begin{equation*}
f_{h}(x) = \int \frac{1}{\prod_{\tau=1}^i f(h_\tau|h_{\tau-1}, c) } f(x,h,c) dc.
\end{equation*}
Finally, in this work the category $c\in\mathscr{C}$ is a discrete variable, so we modify to the discrete case and obtain the final expression:
\begin{equation*}
f_{h}(x) = \sum_{c\in\mathscr{C}} \frac{1}{\prod_{\tau=1}^i f(h_\tau|h_{\tau-1}, c)} f(x,h,c),
\end{equation*}
arriving at \eqref{eq:counterfactual-pdf} that reaches the lemma.

\begin{remark}
In practice, we can consider using the stabilized weight (SW) instead of the IPTW such that the weights are less prone to exploding due to small denominator values. See \cite{hernan2010causal} for more details.
\begin{equation*}
SW(\bar{h},c) = \frac{\prod_{i=1}^{n+1} f(h_i|h_{i-1})}{\prod_{i=1}^{n+1} f(h_i|h_{i-1}, c)}.
\end{equation*}
\end{remark}

% \woody{Replace $h_i$ to $h$. There is a significant difference between these two notations (Check out the figure 1). Similarly in Appendix C. }\cmtH{Thanks. I am still working on this.}

% \begin{equation}
% \begin{split}
% f_{h_i}(x) & = \int f(x|h_i, c) \cdot f(c) dc\\
% & = \int f(x|h_i, c) \cdot \frac{\prod_{\tau=1}^i f(h_\tau|h_{\tau-1}, c)}{\prod_{\tau=1}^i f(h_\tau|h_{\tau-1}, c)} \cdot f(c) dc\\
% & = \int \frac{1}{\prod_{\tau=1}^i f(h_\tau|h_{\tau-1}, c) f(h_0|c)} \left[ f(x|h_i, c) \cdot \prod_{\tau=1}^i f(h_\tau|h_{\tau-1}, c) \cdot f(c) \cdot f(h_0|c)  \right] dc\\
% & = \int \frac{1}{\prod_{\tau=1}^i f(h_\tau|h_{\tau-1}, c)  f(h_0|c)} f(x,h_i,c) dc\\
% & = \sum_{c} \frac{1}{\prod_{\tau=1}^i f(h_\tau|h_{\tau-1}, c)  f(h_0|c)} f(x,h_i,c)
% \end{split}
% \end{equation}
% In the setting of our problem, the initial history embedding $h_0$ starts where there have been no events observed at all, therefore making $h_0$ a constant value. Therefore, $f(h_0|c)$ can be regarded as $1$. This results in the final expression
% \begin{equation}
% f_{h_i}(x) = \sum_{c} \frac{1}{\prod_{\tau=1}^i f(h_\tau|h_{\tau-1}, c)  } f(x,h_i,c)
% \end{equation}

% Therefore, the term $w(\bar{h},c) = \frac{1}{\prod_{\tau=1}^i f(h_\tau|h_{\tau-1}, c)}$ becomes the inverse propensity score derived from this setup. 

\section{Proof of Proposition~\ref{prop:pesudo-objective}}
\label{append:prop-1}
The learning objective function for weighted maximum log-likelihood estimation can be obtained via the definition of expectation.
% The log-likelihood function $\log f_{\theta} (x|h)$ depends on two random variables $x$ and $h$, so we need to take the expectation with respect to the two random variables, giving us the following expression to evaluate:
% \begin{equation*}
% \mathbb{E}_{h}\left[ \mathbb{E}_{x\sim f_{h}}  \left[ \text{log}f_{\theta}(x|h)\right]\right].
% \end{equation*}
We twice expand the expectation with respect to $f(x|h)$ (or $f_{h}(x)$) and the event history variable $h$:
\begin{equation*}
\begin{split}
\mathbb{E}_{h}\left[ \mathbb{E}_{x\sim f_{h}}  \left[ \text{log}f_{\theta}(x|h)\right]\right] &= \mathbb{E}_{h}\left[ \int \text{log}f_{\theta}(x|h)\cdot f_{h}(x)dx\right]\\
&= \int \int \text{log}f_{\theta}(x|h)\cdot f_{h}(x)\cdot f(h)dxdh
\end{split}
\end{equation*}
The distribution of history embedding variable $h$ is typically customized. In this work, we wish to treat each history variable equally, and combined with computational simplicity, we set $f(h) = 1$. Furthermore, using the expression of $f_{h}(x)$ in \eqref{eq:counterfactual-pdf} from Lemma~\ref{lemma:pseudo-prob} while treating $C$ as continuous, we obtain
\begin{equation*}
\begin{split}
\mathbb{E}_{h}\left[ \mathbb{E}_{x\sim f_{h}}  \left[ \text{log}f_{\theta}(x|h)\right]\right] 
&= \int\int \text{log}f_{\theta}(x|h)\cdot \int \frac{1}{\prod_{\tau=1}^i f(h_{\tau}|h_{\tau-1},c)} f(x,h,c) dc dx dh\\
&=  \int\int\int \text{log}f_{\theta}(x|h)\cdot \frac{1}{\prod_{\tau=1}^i f(h_{\tau}|h_{\tau-1},c)} f(x,h,c) dc dx dh.
\end{split}
\end{equation*}
Denote $1/\prod_{\tau=1}^i f(h_{\tau}|h_{\tau-1},c)$ by $w(\bar{h},c)$ and take the expectation with data tuple $(x,h,c)$, we then reach
\begin{equation*}
\begin{split}
\mathbb{E}_{h}\left[ \mathbb{E}_{x\sim f_{h}}  \left[ \text{log}f_{\theta}(x|h)\right]\right] &= \int \int\int \text{log}f_{\theta}(x|h) w(\bar{h},c) f(x,h,c) dh dc dx \\
&= \mathbb{E}_{(x, h, c)}\left[ w(\bar{h},c)\text{log} f_{\theta}(x|h)\right]\\
&\approx \frac{1}{|\mathcal{D}|} \sum_{(x_i^{(k)},\bar{h}_i^{(k)},c)\in\mathcal{D}} w(\bar{h}_i^{(k)},c) \log f_{\theta}(x_i^{(k)}|h_i^{(k)}),\\
&\propto \frac{1}{K} \sum_{(x_i^{(k)},\bar{h}_i^{(k)},c)\in\mathcal{D}} w(\bar{h}_i^{(k)},c) \log f_{\theta}(x_i^{(k)}|h_i^{(k)}),
\end{split}
\end{equation*}
which can be approximated by samples of data tuples $(x_i^{(k)},\bar{h}_i^{(k)},c)$ from the whole set $\mathcal{D}$, where $|\mathcal{D}|$ denotes the number of data tuples, thus arriving at the first half of \eqref{eq:weighted-log-likelihood}.

The set $\mathcal{D}$ of the data tuples $(x_i^{(k)},\bar{h}_i^{(k)},c)$ contains the history embedding trajectory for all $K$ users, so we can restructure them by $k$:
\begin{equation*}
\frac{1}{K}  \sum_{(x_i^{(k)},\bar{h}_i^{(k)},c)\in\mathcal{D}}   w(\bar{h}_i^{(k)},c) \log f_{\theta}(x_i^{(k)}|h_i^{(k)})
= \frac{1}{K}\sum_{k=1}^K \sum_{i=1}^{N_T^{(k)}} w(\bar{h}_i^{(k)},c) \log f_{\theta}(x_i^{(k)}|h_i^{(k)}).
\end{equation*}
Plug in the expressions for the probability density function from \eqref{eq:cond-prob}, we obtain:
\begin{equation*}
\begin{split}
&~\frac{1}{K} \sum_{(x_i^{(k)},\bar{h}_i^{(k)},c)\in\mathcal{D}}   w(\bar{h}_i^{(k)},c) \log f_{\theta}(x_i^{(k)}|h_i^{(k)}) \\
= &~\frac{1}{K} \sum_{k=1}^K \sum_{i=1}^{N_T^{(k)}} w(\bar{h}_i^{(k)},c) \log \bigg(\lambda_{\theta}(x_i|h_i)\exp \left( -\int_{[t_{i-1}, t_i)\times \mathscr{M}} \lambda_{\theta}(x|h_i)dx \right)\bigg)\\
= &~\frac{1}{K} \sum_{k=1}^K \sum_{i=1}^{N_T^{(k)}} w(\bar{h}_i^{(k)},c) \left( \log \lambda_{\theta} (x_i|h_i) - \int_{[t_{i-1}, t_i)\times \mathscr{M}} \lambda_{\theta}(x|h_i)dx  \right).
\end{split}
\end{equation*}
We then plug in the log-likelihood function expression from \eqref{eq:log-likelihood}:
\begin{equation*}
\begin{split}
= &~\frac{1}{K} \sum_{k=1}^K \left(\sum_{i=1}^{N_T^{(k)}} w(\bar{h}_i^{(k)},c) \log \lambda_{\theta} (x_i|h_i) - \sum_{i=1}^{N_T^{(k)}}\int_{[t_{i-1}, t_i)\times \mathscr{M}} w(\bar{h}_i^{(k)},c) \lambda_{\theta}(x|h_i)dx \right)\\
= &~ \left( \frac{1}{K} \sum_{k=1}^K \sum_{i=1}^{N_T^{(k)}} w(\bar{h}_i^{(k)},c) \log \lambda_{\theta} (x_i|h_i)\right) - \left(\frac{1}{K}\sum_{k=1}^K \int_{\mathscr{X}} w(\bar{h}_n^{(k)},c) \lambda_{\theta}(x|h^{(k)})dx\right) \\
= &~\frac{1}{K}\sum_{k=1}^K  \int_{\mathscr X} w(\bar{h}, c) \log  \lambda_\theta(x|h)  d \mathbb N^{(k)}(x) - \frac{1}{K}\sum_{k=1}^K \int_{\mathscr X} w(\bar{h}_n^{(k)}, c) \lambda_\theta(x|h^{(k)}) dx\\
 =&~ \frac{1}{K}\sum_{k=1}^K \left( \int_{\mathscr X} w(\bar{h}, c) \log \big ( \lambda_\theta(x|h(x)) \big) d \mathbb N^{(k)}(x) - \int_{\mathscr X} w(\bar{h}_n^{(k)}, c) \lambda_\theta(x|h^{(k)}(x)) dx \right) ,
\end{split}
\end{equation*}
finally arriving at the latter half of \eqref{eq:weighted-log-likelihood}.

\section{Details for Sensitivity Analysis}
\label{append:sensitivity}
\subsection{Proof of Proposition~\ref{prop:improve}}
Following the definition of $\widehat{w}_\delta$ in \eqref{eq:w_delta}, we can see that when $\delta=1$, the size of the bin is exactly the same as the sample space, with all samples falling into this single bin. Therefore all conditional transition probabilities are 1, \ie, $\widehat{w}_1(\cdot)=1$. On the other hand, when $\delta<1$ where there are more than one bins, the conditional probabilities $\widehat{f}_\delta(\cdot)\le 1$ which leads to $\widehat{w}_\delta \ge 1$. Thus $\ell^*(\widehat{w}_\delta)\ge \ell^*(\widehat{w}_1)$. Additionally, by definition $\ell^*(\widehat{w}_\delta)$ given any $\delta$ will be smaller than the true $\ell^*(w)$ whatever the value of $\delta$ takes. This leads to $\ell^*(\widehat{w}_\delta)\le \ell^*(w)$.

\begin{remark}
This proposition has shown that the introduction of IPTW weights using approximation via discretization in the history embedding space will achieve a non-negative improvement to the log-likelihood function. This indicates the proposed counterfactual framework is guaranteed to improve the performance of the framework when compared to the method without counterfactual weights, regardless of the choice of $\delta$.
\end{remark}

\subsection{Bias and Variance Analysis}
\begin{lemma}[Bias and Variance by binning]
\label{append:lemma-bias}
Let history embedding variables $h_i$ and $h_{i-1}$ are of one dimension, which can be extended to higher dimension. Suppose there are $m$ samples of the data tuples $(h_i,h_{i-1},c)$, and our objective is to estimate the true joint probability function $f(h_i,h_{i-1},c)$ (shorthanded by $f$) utilizing the binning approximation. Denoting the bin size by $\delta$, then the bias and variance of the estimator $\widehat{f}_\delta$ are as follows:
\begin{align*}
\text{Bias}[\widehat{f}_\delta] &= \left( \frac{1}{2} \frac{\partial f}{\partial h_i} + \frac{1}{2} \frac{\partial f}{\partial h_{i-1}} \right) \delta +  \mathcal{O}(\delta^2) + \mathcal{C}\\
\text{Var}[\widehat{f}_\delta] & = \frac{f}{m\delta^2}+\mathcal{O}(\frac{1}{m}),
\end{align*}
where $\mathcal{C}$ is a constant not dependent on $\delta$ or $m$. Since $\delta$ is presumed to be small, higher order terms denoted by $\mathcal{O}(\delta^2)$ are deemed negligible.
\end{lemma}

\begin{remark}
From the lemma, we observe the bias of estimator increases with the bin size $\delta$, suggesting coarser bins lead to less accurate estimate. The variance shows it increases as $\delta$ decreases. This reflects a trade-off between bias and variance, as smaller bin size makes the bias smaller but will increase the variance in the estimator. Additionally, the variance decreases as $m$ increases, which suggests more data tuples will reduce the variance, as expected.
\end{remark}

The bias and the variance of the estimator $\hat{f}$ constructed by binning provide a way to obtain the optimal bin size by minimizing the integrated mean square error (IMSE) which is formulated as
\begin{align*}
\begin{split}
\text{IMSE}(\delta) 
% &= \int \int \text{MSE}\; dh_i dh_{i-1}\\&
= \int \int (\text{Bias}^2 + \text{Var})\; dh_i dh_{i-1}
\end{split}
\end{align*}

\subsection{Proof of Sensitivity Analysis}
\label{append:sensi_proof}
In this analysis, we assume the variables $h_i$ and $h_{i-1}$ are scalars, but it can be easily extended to higher dimension. Looking into the joint probability of the tuple $(h_i,h_{i-1},c)$, since $c$ is a known categorical variable, we may omit it when deriving the probability $p(h_i,h_{i-1})$. Suppose there are $m$ samples of the tuple $(h_i,h_{i-1})$, and we wish to approximate the true joint probability distribution function of $h_i$ and $h_{i-1}$, denoted by $f(h_i,h_{i-1})$, by approximation using binning. Suppose the probability of a sample falling in a particular bin is denoted by $p_m(h_i,h_{i-1})$ and the size of bins in $h_i$ and $h_{i-1}$ are uniform denoted by $\delta$, we can express it as follows:
\begin{align*}
p_m(h_i, h_{i-1}) = \int_{t(h_i)}^{t(h_i)+\delta} \int_{t(h_{i-1})}^{h_{i-1}+\delta} f(x_i,x_{i-1})dx_i dx_{i-1},
\label{eq:p_m}
\end{align*}
where $x_i$ denotes a variable in the domain of $h_i$, and $x_{i-1}$ denotes a variable in the domain of $h_{i-1}$, $t(h_i)$ denotes the lower bound of the bin in $h_i$ and $t(h_i)+\delta$ is then the upper bound, same with $h(t_{i-1})$. 

We expand the expression above using Taylor's series to the first degree which yields
\begin{align*}
p_m(h_i,h_{i-1})=  &\int_{t(h_i)}^{t(h_i)+\delta} \int_{t(h_{i-1})}^{h_{i-1}+\delta} f(h_i,h_{i-1}) + \frac{\partial f(h_i,h_{i-1})}{\partial h_i}(x_i-h_i) \\
& + \frac{\partial f(h_i,h_{i-1})}{\partial h_{i-1}}(x_{i-1}-h_{i-1}) + \mathcal{O}(\delta^2)dx_i dx_{i-1}\\
=& \quad\delta^2 f(h_i,h_{i-1}) + \frac{1}{2} \frac{\partial f}{\partial h_i}\left[\delta^3 - 2\delta^2(x_i-t(h_i)) \right] \\
& + \frac{1}{2} \frac{\partial f}{\partial h_{i-1}}\left[\delta^3 - 2\delta^2(x_i-t(h_{i-1})) \right] +\mathcal{O}(\delta^4),
\end{align*}
where $\mathcal{O}(\cdot)$ indicates higher order terms which are small in magnitude.

Now we denote $\nu_m(h_i,h_{i-1})$ as the number of samples falling in this bin, which therefore follows a binomial distribution (since a sample can either be within or without the bin), denoted by $\mathcal{B}\{ m,p_m(h_i,h_{i-1})\}$. This leads to 
\begin{align*}
\mathbb{E}[\nu_m] &= mp_m,\\
\text{Var}[\nu_m] &= mp_m(1-p_m).
\end{align*}
Given $\nu_m$, we can write out the estimator for the true probability density function $\widehat{f}_\delta$:
\begin{equation*}
\widehat{f}_\delta = \frac{\nu_m}{m \delta^2}.
\end{equation*}
We then thus express the expectation and variance of the estimator $\widehat{f}_\delta$:
\begin{align*}
\mathbb{E}[\widehat{f}_\delta]&= \frac{\mathbb{E}[\nu_m]}{m\delta^2} = \frac{p_m(h_i,h_{i-1})}{\delta^2}\\
&= f(h_i,h_{i-1}) + \frac{1}{2} \frac{\partial f}{\partial h_i}\left[ \delta-2(x_i-t(h_i)) \right]+ \frac{1}{2} \frac{\partial f}{\partial h_{i-1}}\left[ \delta-2(x_{i-1}-t(h_{i-1})) \right] + \mathcal{O}(\delta^2);\\
\text{Bias} &= \frac{1}{2} \frac{\partial f}{\partial h_i}\left[ \delta-2(x_i-t(h_i)) \right]+ \frac{1}{2} \frac{\partial f}{\partial h_{i-1}}\left[ \delta-2(x_{i-1}-t(h_{i-1})) \right] + \mathcal{O}(\delta^2);\\
\text{Var}[\widehat{f}_\delta] & = \frac{\text{Var}[\nu_m]}{m^2\delta^4} = \frac{p_m(1-p_m)}{m\delta^4}\\
&= \frac{1}{m}\frac{1}{\delta^4}\left[ f(h_i,h_{i-1})\delta^2+\mathcal{O}(\delta^4)\right]\left[1-\mathcal{O}(\delta^2) \right]\\
&= \frac{f(h_i,h_{i-1})}{m\delta^2}+\mathcal{O}(\frac{1}{m}).
\end{align*}

Then the mean square error for the estimator can be formulated as follows:
\begin{align*}
\text{MSE} =&\; \text{Bias}^2 + \text{Var}\\
=&\; \frac{f(h_i,h_{i-1})}{m\delta^2} + \mathcal{O}(\frac{1}{m}) + \mathcal{O}(\delta^3) \\
&+\frac{1}{4}(\frac{\partial f}{\partial h_i})^2\delta^2 + (\frac{\partial f}{\partial h_i})(h_i-t(h_i))^2 + \frac{1}{4}(\frac{\partial f}{\partial h_{i-1}})^2\delta^2 + (\frac{\partial f}{\partial h_{i-1}})(h_{i-1}-t(h_{i-1}))^2 \\
& -\delta(\frac{\partial f}{\partial h_i})^2(h_i-t(h_i)) - \delta(\frac{\partial f}{\partial h_{i-1}})^2(h_{i-1}-t(h_{i-1})) +\frac{1}{2} \frac{\partial f}{\partial h_i}\frac{\partial f}{\partial h_{i-1}}\delta^2\\
&+ 2\frac{\partial f}{\partial h_i}\frac{\partial f}{\partial h_{i-1}} (h_i-t(h_i))(h_{i-1}-t(h_{i-1})) - \delta\frac{\partial f}{\partial h_i}\frac{\partial f}{\partial h_{i-1}} (h_{i-1}-t(h_{i-1})) \\
&- \delta \frac{\partial f}{\partial h_i}\frac{\partial f}{\partial h_{i-1}}(h_i-t(h_i)).
\end{align*}

Now we make the assumption here that the number of events in the sequence is long enough, such that we expect the joint distribution of the embedding between two steps $f(h_i,h_{i-1})$ to be symmetric between $h_i$ and $h_{i-1}$ since the transition behavior is expected to converge. Then we can argue that the two partial derivatives are equivalent in magnitude given the symmetry and can be denoted so:
\begin{align*}
\frac{\partial f}{\partial h_i} = \frac{\partial f}{\partial h_{i-1}} = f'.
\end{align*}

Now taking the integral over $(x_i,x_{i-1})$, the integrated mean square error (IMSE) can be found by taking the two-dimensional integral of all the terms in MSE. The first few terms are straightforward to integrate:
\begin{align*}
\int_{t(h_i)}^{t(h_i)+\delta}\int_{t(h_{i-1})}^{t(h_{i-1})+\delta}  \frac{f(h_i,h_{i-1})}{m\delta^2} + \mathcal{O}(\frac{1}{m}) + \mathcal{O}(\delta^3)dh_i dh_{i-1} = \frac{1}{m\delta^2} + \mathcal{O}(\frac{1}{m})+ \mathcal{O}(\delta^5).
\end{align*}
The next four terms can be derived directly:
\begin{align*}
\int\int \frac{1}{4}(\frac{\partial f}{\partial h_i})^2\delta^2 dh_idh_{i-1} & = \frac{1}{4}\delta^3 \int (\frac{\partial f}{\partial h_i})^2 dh_i = \frac{1}{4}\delta^3 \int f^{'^2},\\
\int\int (\frac{\partial f}{\partial h_i})(h_i-t(h_i))^2 dh_idh_{i-1} &= \frac{1}{3} \delta^3 \int f^{'^2} +\mathcal{O}(\delta^3);\\
\int\int \delta(\frac{\partial f}{\partial h_i})^2(h_i-t(h_i)) dh_idh_{i-1} &= \frac{1}{2} \delta^3 \int f^{'^2} + \mathcal{O}(\delta^3);\\
\int\int \frac{1}{2} \frac{\partial f}{\partial h_i}\frac{\partial f}{\partial h_{i-1}}\delta^2 dh_i dh_{i-1} &= \frac{1}{2}\delta^2 \left(\int f'\right)^2 +\mathcal{O}(\delta^3).
\end{align*}
The derivation of the final two terms requires the use of the assumption on the two partial derivatives:
\begin{align*}
2&\frac{\partial f}{\partial h_i}\frac{\partial f}{\partial h_{i-1}} (h_i-t(h_i))(h_{i-1}-t(h_{i-1})) \\
&= 2\int\frac{\partial f}{\partial h_i}(h_i-t(h_i)) \int \frac{\partial f}{\partial h_{i-1}}(h_{i-1}-t(h_{i-1}))dh_{i-1}\\
&= 2\left( \frac{1}{2}\delta^2 \int \frac{\partial f}{\partial h_i}dh_i\right)\left( \frac{1}{2}\delta^2 \int \frac{\partial f}{\partial h_{i-1}}dh_{i-1}\right) + \mathcal{O}(\delta^5)\\
&= \frac{1}{2}\delta^4 \left(\int f'\right)^2+\mathcal{O}(\delta^5).
\end{align*}
The final term can be derived in a similar way:
\begin{align*}
\delta&\frac{\partial f}{\partial h_i}\frac{\partial f}{\partial h_{i-1}} (h_{i-1}-t(h_{i-1})) \\
&= \delta \int\frac{\partial f}{\partial h_i}dh_i \int \frac{\partial f}{\partial dh_{i-1}}(h_{i-1}-t(h_{i-1}))dh_{i-1}\\
&= \delta \int\frac{\partial f}{\partial h_i}dh_i \left( \frac{1}{2}\delta^2 \int\frac{\partial f}{\partial h_{i-1}}dh_{i-1}+\mathcal{O}(\delta^3) \right)\\
&= \frac{1}{2}\delta^3 \int\frac{\partial f}{\partial h_i}dh_i \int\frac{\partial f}{\partial h_{i-1}}dh_{i-1}  + \mathcal{O}(\delta^4)\\
&= \frac{1}{2}\delta^3 \left( \int f'\right)^2 +\mathcal{O}(\delta^4).
\end{align*}
When put together, the IMSE can be expressed as
\begin{align*}
% \text{IMSE} &= \frac{1}{m\delta^2} + \frac{1}{2} \delta^2 \left( \int f' \right)^2 + \mathcal{O}(\delta^3)+(\frac{1}{4}+\frac{1}{3}+\frac{1}{4}+\frac{1}{3}-\frac{1}{2}-\frac{1}{2})\delta^3 \int f^{'^2} - \delta^3 \left( \int f'\right)^2 + \mathcal{O}(\delta^4)\\
\text{IMSE} &= \frac{1}{m\delta^2} + \frac{1}{2} \delta^2 \left( \int f' \right)^2 + \mathcal{O}(\delta^3) + \mathcal{O}(\frac{1}{m}).
\end{align*}
To obtain the best value of $\delta^*$, we set the derivative of IMSE to 0 dropping the higher order terms and solve the equation.
\[
\frac{-2}{m\delta^3} + \delta\left( \int f^{'} \right)^2=0~\Rightarrow~
\delta^* = \left( \frac{2}{\left( \int f^{'} \right)^2}  \right)^{1/4} m^{-1/4}.
\]
Therefore, we have reached the approximate estimator of the optimal bin size $\delta^*$ which minimizes the IMSE caused by bin discretization. This applies when $\int f^{'}\neq 0$. 

\noindent In the event where $\int f^{'}= 0$, the IMSE is formulated differently:
\begin{align*}
\text{IMSE} &= \frac{1}{m\delta^2} +(\frac{1}{4}+\frac{1}{3}+\frac{1}{4}+\frac{1}{3}-\frac{1}{2}-\frac{1}{2})\delta^3 \int f^{'^2} + \mathcal{O}(\delta^4)\\
&= \frac{1}{6} \delta^3 \int f^{'^2} + \mathcal{O}(\delta^4).
\end{align*}
We similarly take the derivative to find the optimal $\delta$.
\[
\frac{-2}{m\delta^3} + \frac{1}{2}\delta^2 \int f^{'^2} = 0~\Rightarrow~
\delta^* = \left( \frac{4}{\int f^{'^2}}  \right)^{1/5} m^{-1/5}.
\]
\begin{remark}
Proposition~\ref{append:opt-delta} indicates the asymptotically optimal choice of bin size depends on both the partial derivative $f'$ and the number of samples $m$. A larger $m$ pushes the optimal choice of $\delta$ smaller. For example, a ten-fold increase in sample numbers $m$ will shrink $\delta^*$ by a factor of 0.56. It also depends on the probability density function $f$ as well. If $f$ is a bivariate normal distribution with zero mean $\mu=0$, isotropic deviation $\sigma=1$, and correlation coefficient $\rho$, the approximated $\delta^*\approx (24\pi^2(1-\rho)(1+\rho)^3/m)^{1/5}$. 
% For detailed proof refer to the appendix.

\noindent Another approach to this can be found in \cite{hahn2022feature}.
\end{remark}

\section{Learning Algorithm}
\label{sec:algorithm}
Calculating the objective function in \eqref{eq:weighted-log-likelihood} requires more attention. The first integral can be directly computed, while the second integral can only be approximated. We utilize a discrete sampling scheme to approximate the second integral. In this work, the event space $\mathscr{X}=[0,T)\times\mathscr{M}$ is constructed on the continuous temporal component and the discrete mark, so only the former is discretized for sampling. The event space $\mathscr{X}=[0,T)\times\mathscr{M}$ is constructed on the continuous temporal space and the discrete mark space, so we need only to discretize the temporal space. Suppose the temporal space can be evenly partitioned into $s$ sections by a set of discrete grid points $\bar{t}_0, \bar{t}_1,\dots,\bar{t}_s$, then we have
\begin{equation}
\int_{\mathscr{X}} w(\bar{h}_n^{(k)},c)\lambda_{\theta}(x|h^{(k)}(x))dx
\approx \frac{1}{|\mathscr{M}|}\sum_{t=\bar{t}_0}^{\bar{t}_s} \sum_{m\in\mathscr{M}} w(\bar{h}_n^{(k)},c)\lambda_{\theta}(t,m|h^{(k)}) \Delta t,
\label{eq:discrete_approx}
\end{equation}
where $\Delta t = \bar{t}_1-\bar{t}_0$ is the size of the grid element.

\begin{algorithm}[!t]
\caption{Joint Learning Algorithm}
\label{alg:joint_opt}
\KwData{Event sequences of $K$ users: $\mathcal{D} = \{x^{(1)_{}}, x^{(2)},\dots,x^{(K)}\}$;\\
\quad\quad\quad \hspace{0.000005in} Number of events of $K$ users: $N_T^{(1)}, N_T^{(2)}, \dots, N_T^{(K)}$;\\
% $\{ x_1^{(k)}, x_2^{(k)}, \dots,x_{N_T^{(k)}}^{(k)} \}_{k=1}^K$;\\
\quad\quad\quad \hspace{0.000005in} Category of $K$ users: $c^{(1)}, c^{(2)},\dots, c^{(K)}$.
%\quad\quad\quad \hspace{0.000005in} Bin size: $\delta$; \quad Number of epochs for IPTW update: $\eta$;\\
%\quad\quad\quad \hspace{0.000005in} Number of training epochs: $E$; \quad Number of batches: $B$.
}
\textbf{Parameters:} Number of training epochs: $E$; \, Number of batches: $B$; \, Hyperparameters: $\delta$, $\eta$.\\
\textbf{Initialization:} IPTW values $w_{\delta} (\bar{h},c) =  1$, $\forall \bar{h}, c$.\\
\For{$e = 1:E$}{
    $\mathcal{G} \leftarrow \emptyset$,
    $\mathcal{F} \leftarrow \emptyset$;\\
    \For{$b = 1:B$}{
        Randomly select $\kappa$ user event sequences $\mathcal{D}_{b} \coloneqq \{x^{(k)}\} \subset \mathcal{D}$;\\
            \For{$k = 1:\kappa$}{
                \For{$i = 1:N_T^{(k)}$}{
                    $h^{(k)}_{i} \leftarrow \phi(x_i^{(k)}, h^{(k)}_{i-1})$ for all $x_i\in x^{(k)}$ via \eqref{eq:history-var};\\
                    $\mathcal{G} \leftarrow \mathcal{G} \cup \{(h^{(k)}_{i}, h^{(k)}_{i-1}, c^{(k)})\}$;\\
                    $\bar{h}_i^{(k)} \leftarrow (h_0^{(k)}, \dots, h_i^{(k)})$;\\
                    $\mathcal{F} \leftarrow \mathcal{F} \cup \{ (x_i^{(k)}, \bar{h}_i^{(k)}, c^{(k)})\}$;
                }
            }
    }
    $S \leftarrow 0$;\\
    \For{$(x_i^{(k)}, \bar{h}_i^{(k)}, c^{(k)})$ in $\mathcal{F}$}{
        Obtain $f(x_i^{(k)}|h_i^{(k)})$ from $\lambda(x_i^{(k)}|h^{(k)}(x))$ via \eqref{eq:cond-prob} \& \eqref{eq:nn-lam} ;\\
        $S \leftarrow S + w_\delta(\bar{h}_i^{(k)}, c^{(k)}) \log f(x_i^{(k)}|h^{(k)}(x))$;
    }
    $\theta \leftarrow \text{arg}\max_\theta S$ in \eqref{eq:weighted-log-likelihood};\\
    \If{$e \sslash \eta=0$}{
        $\mathcal{L} \leftarrow \{ (u,v,r)|1\leq u, \, v\leq 1/\delta, \, 1\leq R, \, u,v,r\in\mathbb{Z}\}$;\\
        \For{$(u,v,r)\in\mathcal{L}$}{
            % \For{$v=1:1/\delta$}{
            %     \For{$r=1:R$}{
                    \small $p(u,v,r) \leftarrow \frac{\sum_{\mathcal{G}} \mathbf{1}[(h_i^{(k)}, h_{i-1}^{(k)}, c^{(k)}) \in \mathscr{B}(u, v, r)]}{  \sum_{k=1}^K N_T^{(k)}}$ via (10);
            %     }
            % }
        }
        \For{$(u,v,r)\in\mathcal{L}$}{
            % \For{$v=1:1/\delta$}{
            %     \For{$r=1:R$}{
                    \small $f_{\delta}(\frac{u-1}{\delta}\leq h_i<\frac{u}{\delta}|\frac{v-1}{\delta}\leq h_{i-1}<\frac{v}{\delta}, c=r) \leftarrow \frac{p(u,v,r)}{\sum_u p(u,v,r)}$ in \eqref{eq:prob_approx};
            %     }
            % }
        }
        \For{$k=1:K$}{
            \For{$i=1:N_T^{(k)}$}{
                $\bar{h}_i^{(k)} \leftarrow (h_0^{(k)}, \dots, h_i^{(k)})$;\\
                $w_\delta(\bar{h}_i^{(k)}, c^{(k)}) \leftarrow 1/\prod_{i'=1}^{i} f_\delta(h_{i'}^{(k)}|h_{i'-1}^{(k)}, c^{(k)})$ via \eqref{eq:w_delta};
            }
        }
    
        % \small $p(u, u', r) \leftarrow\left(\sum_{k=1}^K \sum_{j=1}^{N_T^{(k)}} \mathbbm{1}[(h_j^{(k)}, h_{j-1}^{(k)}, c^{(k)}) \in \mathscr{B}(u, u', r)]\right)/\sum_{k=1}^K N_T^{(k)}$\normalsize in \eqref{eq:bin_prob};\\
        % $f_\delta(h_i|h_{i-1},c)\leftarrow \frac{\sum_{(u,u',r)} p(u, u', r) \mathbbm{1}[(h_i, h_{i-1}, c) \in \mathscr{B}(u, u', r)]  }{\sum_{u=1}^{1/\delta} p(u, u', r) \mathbbm{1}[h_i \in \mathscr{H}_u]}$ in \eqref{eq:prob_approx};\\
        % $w_\delta(\bar{h},c) \leftarrow 1/\prod_i f_\delta(h_i|h_{i-1},c)$ in \eqref{eq:w_delta}.\\
    }
}
% = \frac{1}{\prod_{i=1}^{n+1} f(h_i|h_{i-1}, c)}
\KwResult{Optimal IPTW values $\widehat{w}_\delta(\bar{h},c)$; Optimal model parameters $\hat{\theta}$.}
\end{algorithm}

\section{Inference for Prediction}
\label{sec:inference}
Inference from the counterfactual point process model conducts event prediction for new users. The predictions are obtained given the past event sequences of new users without their category information.
% After the model parameters $\theta$ and the IPTW values $\widehat{w}_\delta(\bar{h},c)$ are obtained from the joint learning algorithm, we utilize the fitted model to carry out event prediction for new users. The predictions can be carried out given the past event sequences of new users without their category information. 
To carry out inference, we need to obtain the counterfactual intensity $\lambda_h(x)$, which is approximated by the conditional intensity function $\lambda_{\theta}(x|h)$. We then proceed to obtain the probability density function from the conditional intensity following \eqref{eq:cond-prob}:
% Given an arbitrary new user with their event history, we obtain the fitted counterfactual intensity function $\lambda_{\theta}(x|h)$ and the corresponding probability density function following \eqref{eq:cond-prob}:
\begin{equation*}
f_{\theta}(x|h) = \lambda_{\theta}(x|h) \cdot \exp \left( -\int_{[t_n,t(x))\times\mathscr{M}} \lambda_{\theta}(u|h) du \right).
\label{eq:pdf_pred}
\end{equation*}
% After the joint learning algorithm is carried out to obtain estimates for the point process model parameters $\hat{\theta}$ and the IPTW weights $\hat{w}(\bar{h}|c)$, we can fit the model using these optimized variables. It can then be utilized to carry out event prediction for new users without their cluster information. 
% With the TPP model trained and the model parameters estimated, it is finally utilized to carry out predictions for the next event for any arbitrary user given their event history. Using a marked temporal point process model in general, we can obtain the next-step prediction from the fitted intensity function. Given the fitted conditional intensity function $\lambda^* (t,m)$, we can obtain the conditional probability function following \eqref{eq:cdf}.
% \begin{equation}
% f^*(t,m) = \lambda^*(t,m) \cdot \text{exp}\left(-\int_{t_{n}<t}^t \lambda^*(\tau,m)d\tau\right),
% \label{eq:cdf_pred}
% \end{equation}
% where $t_n$ denotes the last event before time $t$. 
Subsequently, the prediction for both the next upcoming event $\hat{x}=( \hat{t}, \hat{m})$ can be analytically calculated from above:
\begin{equation*}
\hat{x} = \int_{[t_n,t(x))\times\mathscr{M}} x f_\theta(x|h)dx.
\label{eq:pred_cont}
\end{equation*}
When the mark variable $m$ is discrete, it can be further split to time and mark predictions separately as follows:
\begin{align*}
\begin{split}
\hat{t} &= \int_{t_{n}}^{\infty} \tau \sum_{m\in\mathscr{M}}  
f_{\theta}(\tau,m|h) d\tau,\\
\hat{m} &= \text{arg} \max_{m\in\mathscr{M}} \int_{t_n}^{\infty} f_{\theta}(\tau,m|h) d\tau.
\end{split}
\label{eq:pred_disc}
\end{align*}
The integrals in the expressions above can be estimated via Monte Carlo sampling from the domain. In addition to MC sampling, one can also elect to draw multiple samples via simulation and then take the average to obtain the event prediction. It is typically conducted by using the thinning algorithm \citep{ogata1998space}, which offer more insight on uncertainty of the temporal prediction. See \cite{reinhart2018review} for more details regarding prediction and thinning.

% In practice, one can also elect to draw multiple samples from simulations using the thinning algorithm by \cite{ogata1998space}. Drawing multiple samples instead of direct calculation towards a single point estimate may offer more insights into the uncertainty of the time prediction, as well as provide several likely mark predictions. 

\section{Histogram of \texorpdfstring{$h$}{h} Samples and Conditional Transition Matrix}
When the number of bins $1/\delta$ changes, the histogram of $h$ and the conditional transition probabilities will change accordingly. We visualize them when $1/\delta=5$ and $1/\delta=10$ respectively, in addition to the case when $1/\delta=20$ in Fig.~\ref{fig:cond-trans-prob}. These plots illustrate the change in the conditional transition probabilities as $\delta$ is altered.
\begin{figure}[H]
\centering
% \begin{subfigure}[b]{.24\linewidth}
% \includegraphics[width=\linewidth]{imgs/Hist_R_5.png}
% \caption{$f(h_i|c)$}
% \end{subfigure}
% \hfill
\begin{subfigure}[b]{.32\linewidth}
\includegraphics[width=\linewidth]{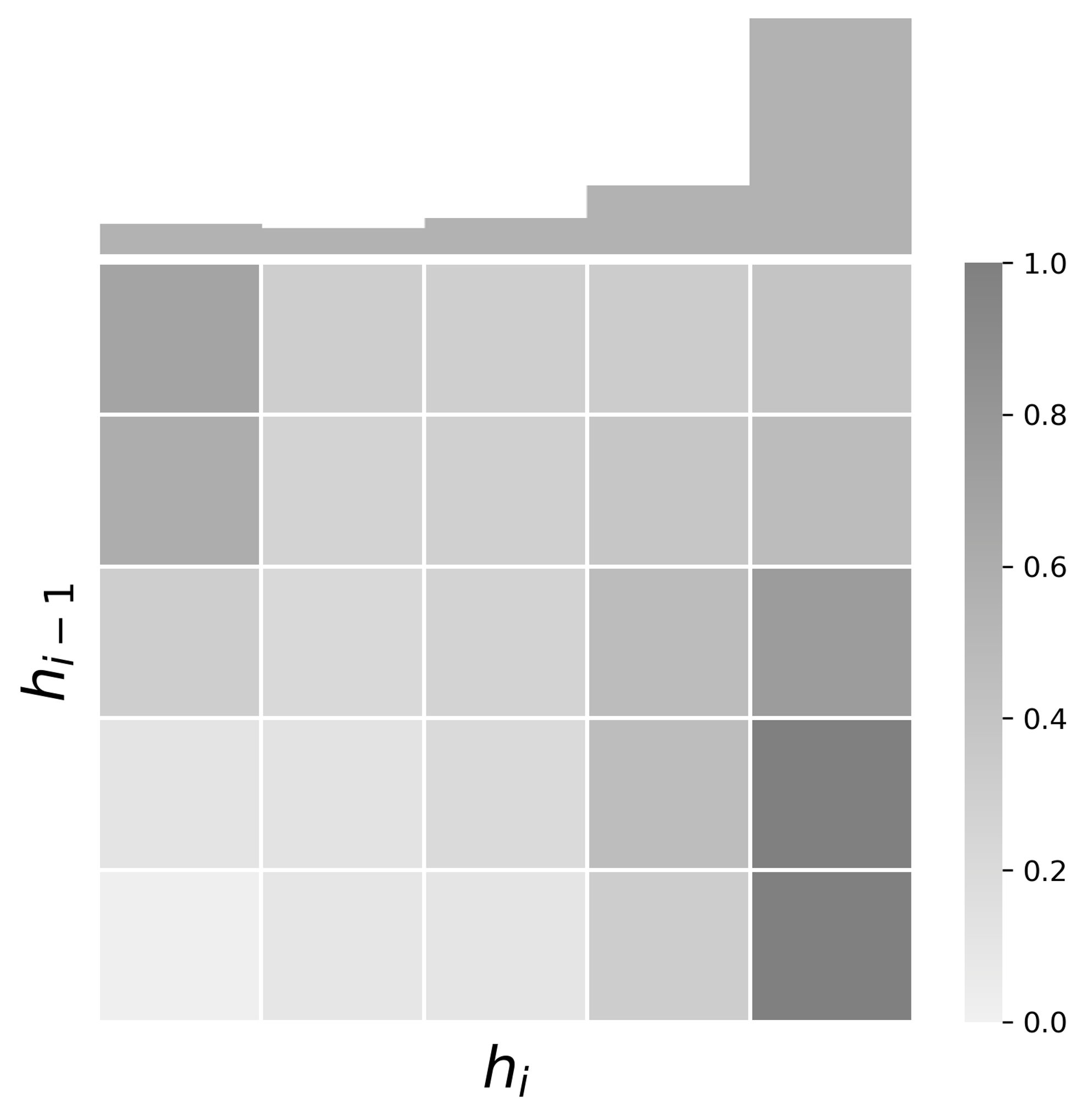}
\caption{$f(h_i|h_{i-1}, c=1)$}
\end{subfigure}
\hfill
\begin{subfigure}[b]{.32\linewidth}
\includegraphics[width=\linewidth]{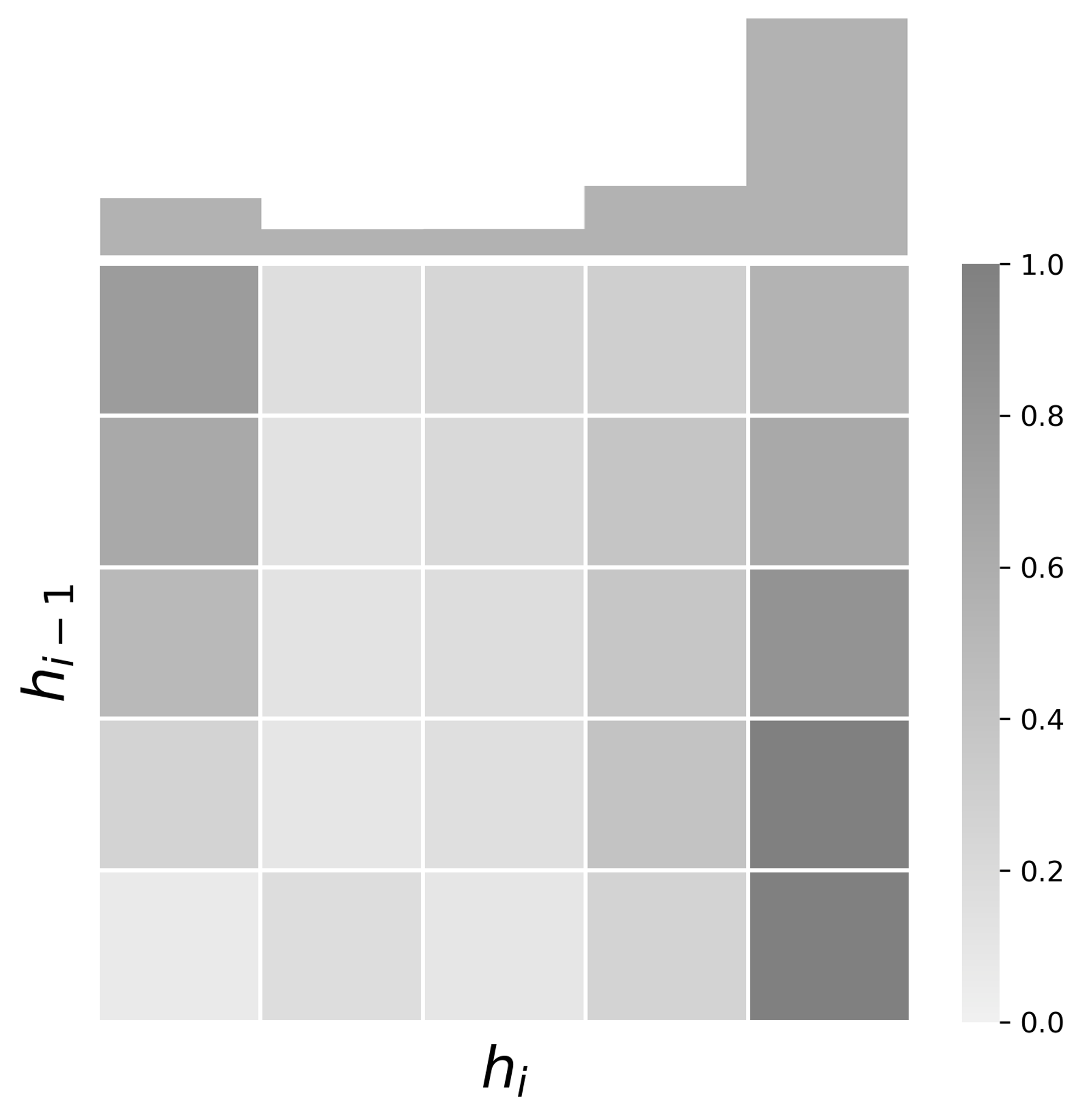}
\caption{$f(h_i|h_{i-1}, c=2)$}
\end{subfigure}
\hfill
\begin{subfigure}[b]{.32\linewidth}
\includegraphics[width=\linewidth]{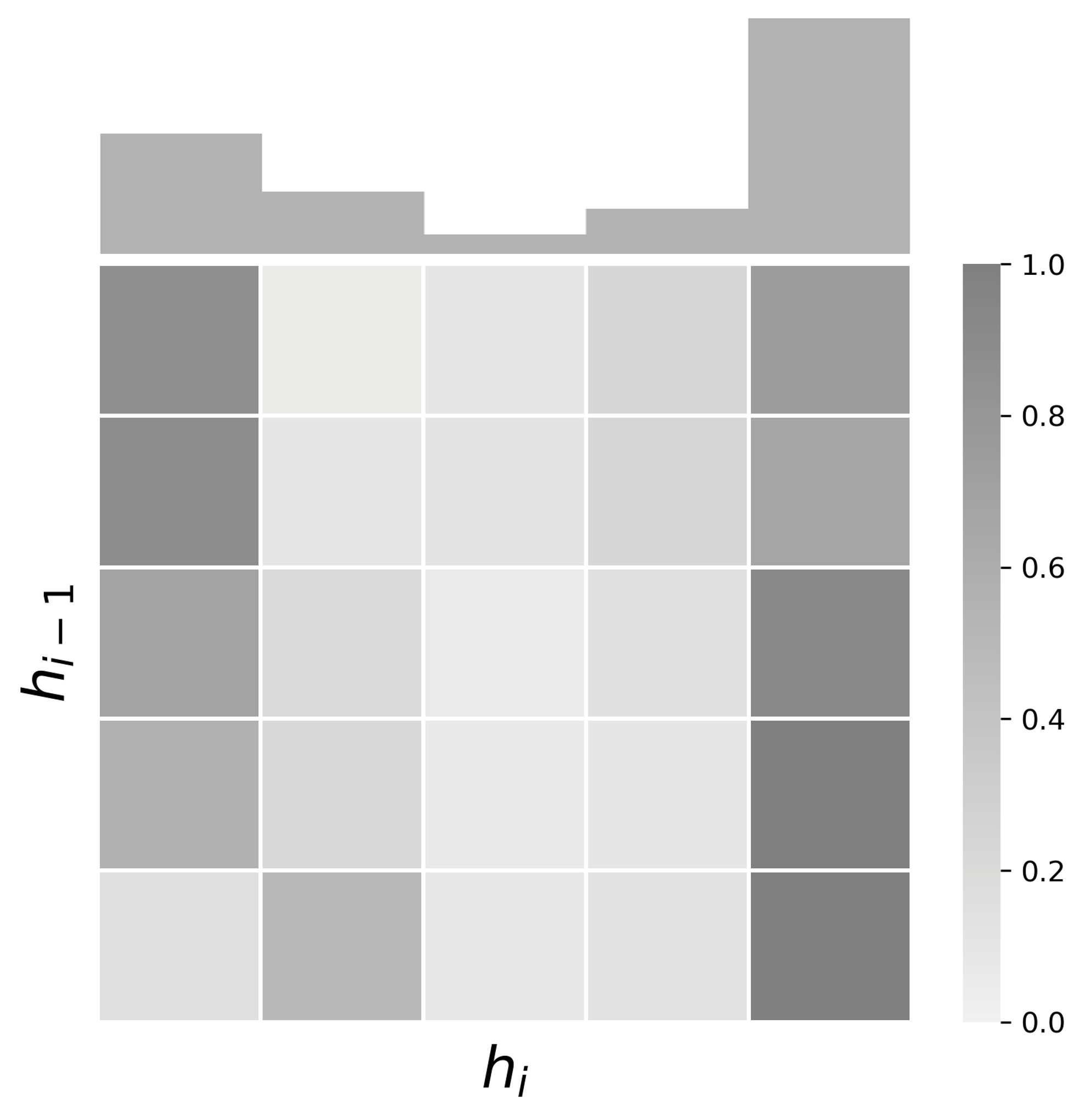}
\caption{$f(h_i|h_{i-1}, c=3)$}
\end{subfigure}
\caption{Histogram of $h$ samples from events in three categories and the corresponding conditional transition matrix when $1/\delta=5$.}

% \begin{subfigure}[b]{.24\linewidth}
% \includegraphics[width=\linewidth]{imgs/Hist_R_10.png}
% \caption{$f(h_i|c)$}
% \end{subfigure}
% \hfill
\begin{subfigure}[b]{.32\linewidth}
\includegraphics[width=\linewidth]{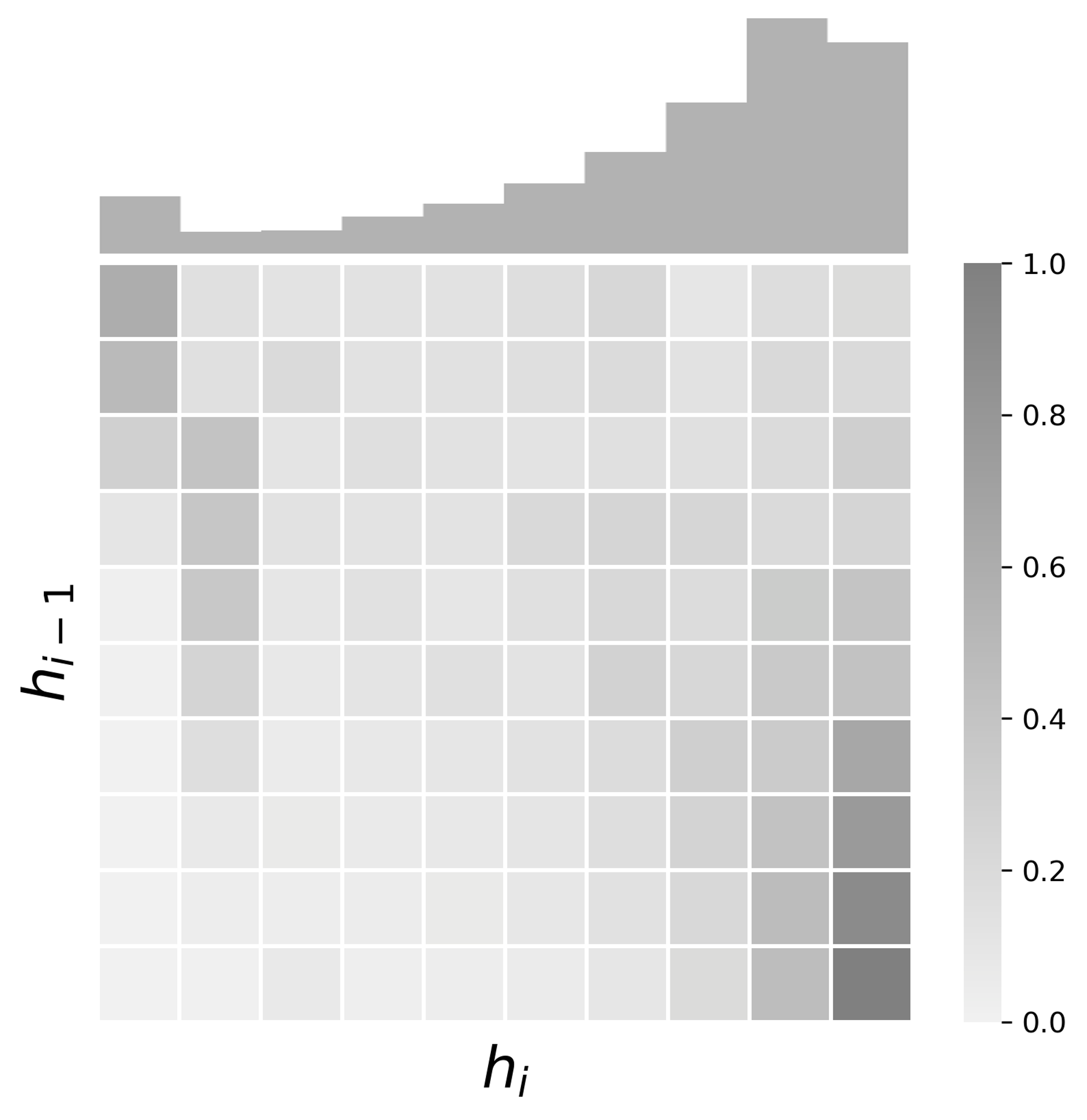}
\caption{$f(h_i|h_{i-1}, c=1)$}
\end{subfigure}
\hfill
\begin{subfigure}[b]{.32\linewidth}
\includegraphics[width=\linewidth]{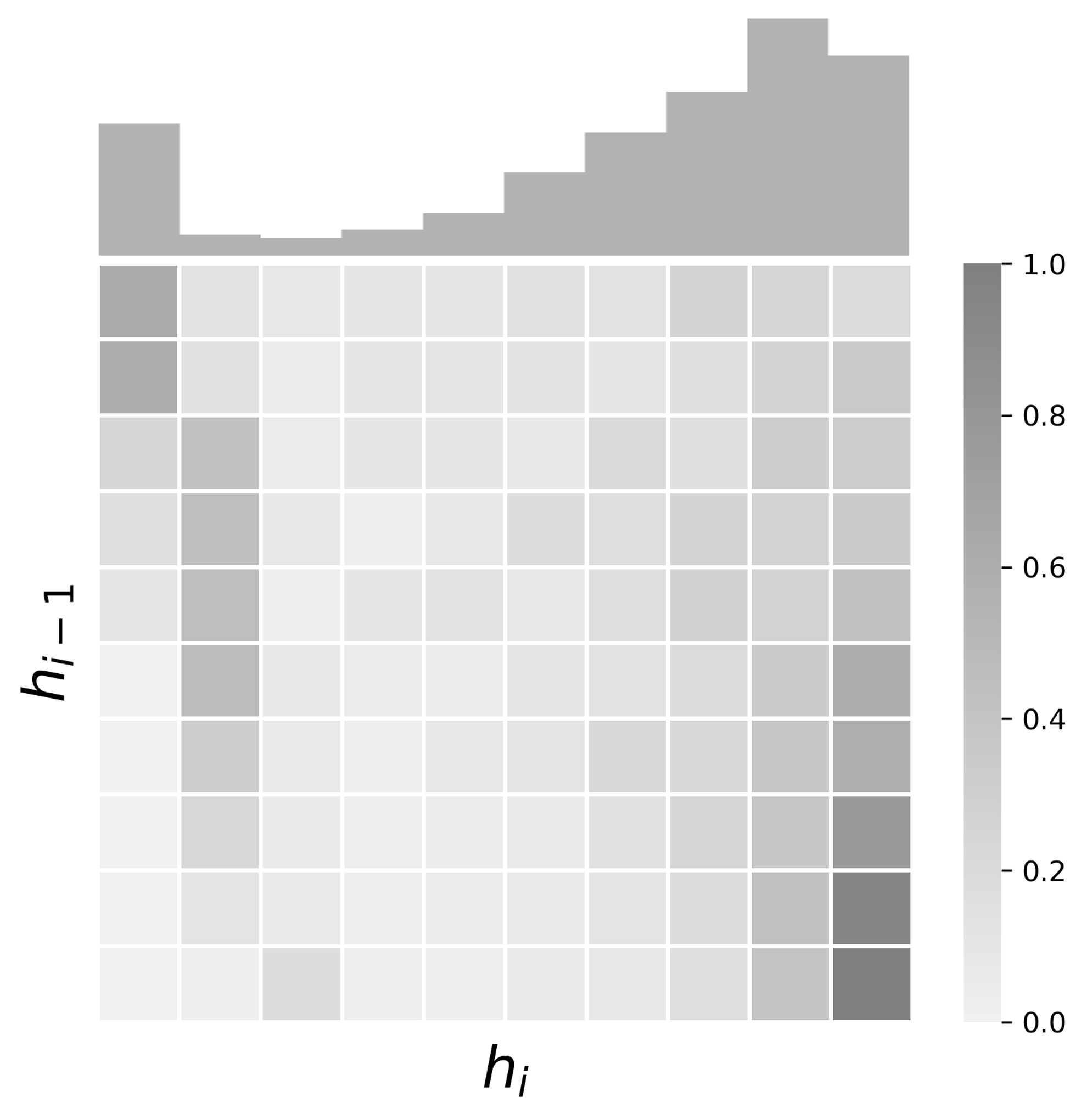}
\caption{$f(h_i|h_{i-1}, c=2)$}
\end{subfigure}
\hfill
\begin{subfigure}[b]{.32\linewidth}
\includegraphics[width=\linewidth]{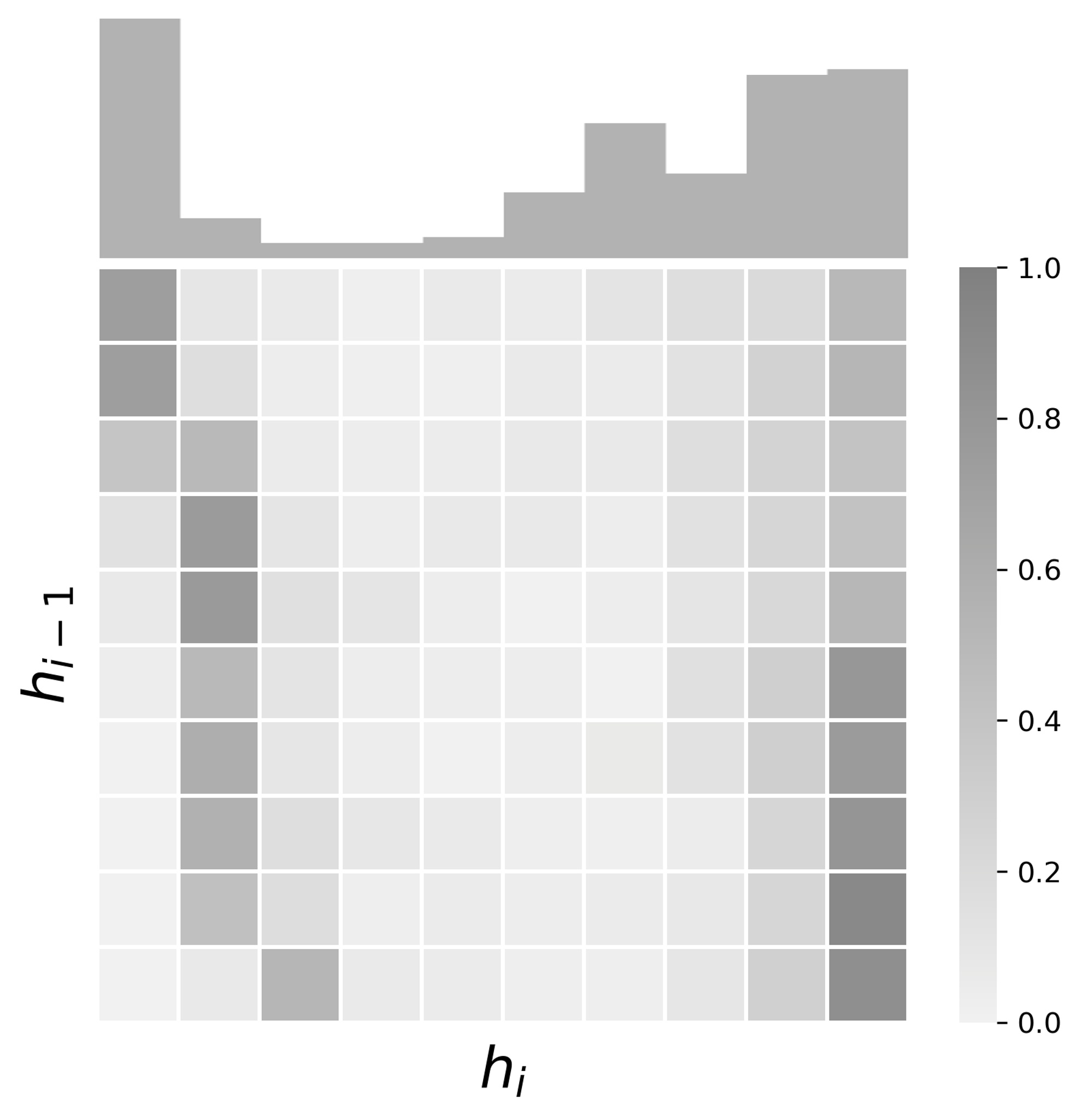}
\caption{$f(h_i|h_{i-1}, c=3)$}
\end{subfigure}
\caption{Histogram of $h$ samples from events in three categories and the corresponding conditional transition matrix when $1/\delta=10$.}

% \begin{subfigure}[b]{.24\linewidth}
% \includegraphics[width=\linewidth]{imgs/Hist_R_20.png}
% \caption{$f(h_i|c)$}
% \end{subfigure}
% \hfill
% \begin{subfigure}[b]{.24\linewidth}
% \includegraphics[width=\linewidth]{imgs/Cluster_1_R_20.png}
% \caption{$f(h_i|h_{i-1}, c=1)$}
% \end{subfigure}
% \begin{subfigure}[b]{.24\linewidth}
% \includegraphics[width=\linewidth]{imgs/Cluster_2_R_20.png}
% \caption{$f(h_i|h_{i-1}, c=2)$}
% \end{subfigure}
% \hfill
% \begin{subfigure}[b]{.24\linewidth}
% \includegraphics[width=\linewidth]{imgs/Cluster_3_R_20.png}
% \caption{$f(h_i|h_{i-1}, c=3)$}
% \end{subfigure}
% \caption{Histogram of $h$ samples from events in three categories and the corresponding conditional transition matrix when $R=20$. \woody{Change color and plot histogram separately. Remove number ticks. }
% }
\label{img:R_study_hist}
\end{figure}

\section{Details for Simulation Study}
\label{sec:simulation-detail}
We investigate a number of synthetic experiments in this study, which we summarize first below and then expand for details.

\begin{enumerate}
    \item Three categories each generate 400 event sequences for the training set, and an additional 100 for the testing one. Sequences within the same category is generated by a temporal Hawkes point process model with exponential triggering function and parameters fixed. Hidden embedding dimension is set to $q=1$ and the number of bins $1/\delta=25$, number of epochs $\eta=5$.
    % Therefore, there are three temporal Hawkes processes used for sequence generation, and the sequence data are evenly divided between them. It indicates the user behavior within the same category is the same, but different across categories, which is commonplace in reality. We set the dimension of the hidden embedding to $p=1$ and the 1D embedding space partitioned into $U=25$ bins to approximate the conditional transition probabilities for the proposed counterfactual methods.
    \item We keep the same setting for the Hawkes models of the three categories, but composition of event sequences from each category is no longer balanced. There are now 200 sequences generated by the first Hawkes model, 400 by the second, and 600 by the third one, thereby creating an unbalanced dataset between three categories. 
    % The composition of number of event sequences for each category is usually unbalanced in the real world. So in the second experiment, we still simulate the sequences from three \texttt{Exp Hawkes} models for the three categories, but this time the composition in the training set is altered. There are now 200 sequences generated by the first Hawkes model, 400 by the second, and 600 by the third one, thereby creating an unbalanced dataset between three categories. 
    \item Composition is further changed. Now 600 sequences are now generated by the first Hawkes model, 400 by the second, and 200 by the third one. This is another unbalanced training set.
    \item Three Hawkes processes for the three categories with parameters no longer fixed. The 1,200 sequences are evenly divided between categories, but the parameter values are drawn from fixed random variables when sampling for each sequence. Event sequences in each category observe a different yet still similar behavior.
    
    % We have so far assumed the user behavior within the same category is the same by utilizing the same Hawkes process model for sequence simulation, which is an idealized version of reality. In fact, it is more common where user behavior in the same category is similar to each other, but not identical. Therefore, in this experiment, we still utilize three Hawkes processes for the three categories, but their model parameters for the conditional intensity are no longer fixed. The 1,200 event sequences are evenly divided between the three Hawkes models, but the model parameters for each are now specified by a prescribed distribution instead of fixed constants. Sample parameter values are then drawn for each sequence when simulating the events for users. The event sequences in each category observe a different, yet still similar, behavior, creating a more complex and realistic synthetic dataset.
    \item We utilize different families of point process models for the three categories. For the first and the second categories, we still utilize two exponential Hawkes process with different parameter. For the third category, we elect to use a Neural Hawkes model. Each category still contributes 400 training and 100 testing sequences.
    
    % So far we have utilized the same family of point process models, \ie, the Hawkes point process in generating the event sequences. Things can be more complex in reality, therefore in this experiment, we utilize different families of PP models for the three categories instead. For the first and the second categories, we still utilzie two exponential Hawkes process with different parameter. For the third category, we elect to use a Neural Hawkes model. Each category still contributes 400 training and 100 testing sequences.
    \item We look into more complex event data with an additional mark variable, \ie the marked point process. We append an additional discrete mark variable to the event sequences generated. We utilize three marked-temporal Hawkes processes for simulation, each contributing the same number of sequences.
    
    % Usually events hold more information then simply the time things happen. Point process models incorporate further information using an additional mark variable, named as marked point process. In this experiment, we append an additional discrete mark variable to the event sequences generated. We utilize three marked-temporal Hawkes processes for simulation, each contributing the same number of sequences.
\end{enumerate}

\paragraph{Synthetic Experiment 1-3}
In this scenario, the event sequences are generated from three independent Hawkes point processes with the exponential kernel function, each corresponding to sequences within a user cluster. Temporal Hawkes point processes without marks utilize the following conditional intensity function:
\begin{equation*}
\lambda_g(t|\mathcal{H}_t) = \mu + \int_0^t g(t-u)d\mathbb{N}(u) = \mu + \sum_{i:t_i<t}g(t-t_i),
\label{eq:hawkes_temp}
\end{equation*}
where $\mu$ is the constant background rate variable, and $g(\cdot)$ is the triggering function that determines the form of self-excitation \citep{reinhart2018review}. Since $\lambda(t|\mathcal{H}_t)$ is non-negative, $g(u)\geq 0$ for $u\geq 0$ and $g(u)=0$ otherwise. Here we designate the triggering function to take the exponential form
\begin{equation*}
g(t-t_i) = \alpha \cdot \text{exp}\left(-\beta (t-t_i)\right).
\end{equation*}
In this scenario, we fix $\mu=0.1$, $\alpha=1$, and alter $\beta$ to generate event sequences of different patterns. For each of the three temporal Hawkes point process models we use to generate the event sequences, we set $\beta=0.5$, 1.0, and 1.5 respectively. Thereby we generate 400 sequences from each of the three PPs in the time range of $T\in[0, 100]$, and the training set contains 1,200 event sequences coming from the three clusters, divided evenly between them such that each cluster contains 400 event sequences. For the testing dataset, we obtained 100 user sequences from each Hawkes PP instead with cluster information removed, a total of 300 sequences. Therefore, we get 111,499 events in the training set, and 15,974 in the testing set, which is sufficiently large. For evaluation, their cluster information is masked so the testing sequences are regarded as coming from ''new'' users. We construct our models with weights upon NH and RMTPP models, where the dimension of hidden variables is set to $q=1$ given the relatively simple mechanism. We partition this 1D embedding space into $S=25$ bins of equal size to obtain the conditional history transition probabilities when deriving the IPTW values for our framework.

Following the simulation setting, we obtain the MAE values for each of the 100 testing sequences drawn from their corresponding exponential Hawkes PP. We look into the comparison between the Neural Hawkes model and the Weighted NH with the proposed reweighting scheme applied more carefully. From Table~1, we can see that the Weighted NH does better than NH overall in terms of MAE, by roughly 11\%. The Weighted RMTPP does better as well, by 3\%.

Additionally, we study the impact on the proposed framework and the baseline methods by different testing data compositions from the three clusters in the testing set. This corresponds to the situation where the composition of event sequences in the training set is different from that in the testing set. We investigate two of such scenarios, denoted by Exp. 2 and 3, respectively. Both of them still entail 1,200 training sequences split evenly from the same three exponential Hawkes PP model, but the composition of their testing set changes. For Synthetic Exp. 2, the numbers of testing event sequences generated by the three Hawkes PP models are 50, 100, and 150 respectively. While for Synthetic Exp. 3, the numbers of testing sequences are 150, 100, and 50 instead. Configurations during training remain the same. The quantitative results are also included in Table~1. The results suggest that in both cases, the proposed methods better predict performance than the baseline models. Specifically, for the comparison between NH and the Weighted NH methods, the improvement in the two scenarios is by 5\% and 10\%, respectively. The improvement over the RMTPP model using our method is 2.4\% and 3.1\%, respectively. This demonstrates that our proposed framework indeed can carry out predictions more accurately for users with event patterns less observed, even when the data composition is imbalanced.

\paragraph{Synthetic Experiment 4}
In the previous section, we look into the case where event sequences in the same cluster are generated from the same PP model. However, in reality, things are usually less ideal. The pattern of events within the same cluster is likely to be similar, but rarely identical. In this scenario, we look into the situation where the event sequences within the same cluster are no longer drawn from the same fixed PP model, yet still they are relatively similar to each other in the cluster. 

% The base scenario utilizes three distinctive Hawkes point process models to generate event sequences for each cluster, where each PP model has fixed parameters. However, in reality things are usually less ideal. To this end, we explore in this section a more complex situation where sequences from each cluster is no longer generated from a single fixed PP model.

Instead of a Hawkes process with a constant fixed $\beta$ for each cluster, we now consider $\beta$ as a random variable, which is sampled randomly whenever an event sequence is simulated. This way, the pattern of event sequences within the cluster will be similar, but no longer identical, creating a more complex and realistic dataset for the cluster. To differentiate between the three clusters, in this case, we define the random variable $\beta$ for each cluster as follows:
\begin{align*}
C_1:&\quad \beta\in\mathcal{U}[0.4,0.6];\\
C_2:&\quad \beta\in\mathcal{U}[0.9,1.1];\\
C_3:&\quad \beta\in\mathcal{U}[1.4,1.6],
\end{align*}
which indicates the $\beta$ are uniform random variables in their respective ranges for each cluster. This more complex event sequence generation for each cluster applies to both the 1,200 training sequences and the 300 testing sequences. The other configurations stay the same, and we can obtain the MAE results in Table~1. We can see that between the MAE from NH and Weighted NH, the latter has a lower MAE by 12\%, which suggests our proposed model works better than baselines in this scenario. For RMTPP, the improvement is about 9\%.

% To differentiate between three clusters, the distributions of $\beta$ for each cluster are shown below:
% \begin{itemize}
% \item $C_1$: $\beta\in[0.4,0.6]$;
% \item $C_2$: $\beta\in[0.9,1.1]$;
% \item $C_3$: $\beta\in[1.4,1.6]$.
% \end{itemize}
% This more complex data generation for each cluster applies to both the 1,200 training sequences and the 300 testing sequences. To evaluate the performance of the proposed method and the NH model, we carry out this experiment from generation to final prediction of the conditional intensity function.

\paragraph{Synthetic Experiment 5}
So far the PP models used in event sequence generation are all exponential Hawkes PP. In this scenario, we study the situation where event sequences are drawn from different distribution families. For the first two clusters in this experiment, we continue to draw event sequences from exponential Hawkes PP, with $\beta=0.5$ and 1, respectively. However, for the third cluster in this case, we utilize an individual Neural Hawkes PP model with parameters randomly set to generate both training and testing event sequences. We set the hidden variable dimension to two when generating sequences from the NH model, which produces 400 training sequences and an additional 100 testing sequences for the third cluster.

Therefore, the total number of event sequences drawn from the three clusters remains the same. When we train our models and the baselines, the dimension of hidden variables is set to $q=3$, higher than the dimension used in generation. We partition this 3D embedding space into $1/\delta^3=5^3=125$ bins of equal size for the conditional history transition. The MAE results for all the methods are logged again in Table~1. We see that for both NH and RMTPP models, our proposed framework improves their performance in the weighted counterpart by 3\% and 10\% respectively. It demonstrates our framework can well handle the situation where the event sequence pattern is no longer identical within the same cluster.

% we obtain user event sequences from five different clusters. For each cluster, we utilize an individual NH model with parameters randomly configured to generate both training and testing sequences. We set the hidden variable dimension to 3 when generating sequences from the NH models. Here we obtain 300 sequences from each NH model, totaling 1,500 sequences for the training set. Similarly, 100 testing sequences are drawn from each cluster to form a testing set of 500.

% For the evaluation process, we set the dimension of the hidden embedding for training to be $p=4$ so that both models being studied are able to capture the intricate relationship between events. For computational efficiency, we partition $h$ into $S=81$ even bins, or three elements in each dimension.

\paragraph{Synthetic Experiment 6}
In this synthetic experiment, we study the case where event sequence data contain an additional categorical mark variable on top of the time component. Marked temporal event data are common in real applications, making this study relevant. In this scenario, the conditional intensity function is constructed as follows:
\begin{equation}
\lambda(t, m|\mathcal{H}_t) = \lambda_g(t|\mathcal{H}_t)f(m|t),
\label{eq:mtpp}
\end{equation}
where $\lambda_g(t|\mathcal{H}_t)$ is the conditional intensity for the temporal Hawkes PP as in \eqref{eq:hawkes_temp}, and the mark component is defined by $f(m|t)$ dependent on time only. It can be described using probability mass functions. 
\begin{align*}
    f(m = 1|t)=&
    \begin{cases}
        0.8 ,& \text{if } t \leq 40,\\
        0,              & \text{otherwise},
    \end{cases}\\
    f(m = 2|t)=& 
    \begin{cases}
        0.8 ,& \text{if } 40 < t \leq 80,\\
        0,              & \text{otherwise},
    \end{cases}\\
    f(m = 0|t)=& 
    \begin{cases}
        0.2 ,& \text{if }  t \leq 80,\\
        1,              & \text{otherwise}.
    \end{cases}
\end{align*}
Therefore, the event sequences in this scenario are still first drawn from exponential Hawkes PPs and then appended with the additional mark, which can only take values between $(0,1,2)$. Given the introduction of an additional mark variable, we again set the dimension of the hidden variable to $q=3$, and the number of bins to $S=125$. The numerical results are appended in Table~1
. For a more detailed breakdown on MAE values across different categories in the testing set, the results are illustrated in Fig.~4.

From the results, we can see our weighted methods generally perform better than the baselines. Comparing NH and Weighted NH, we see the MAE from the proposed weighted method is lower by 4\%. For RMTPP, the improvement is 2.3\%. It indicates our proposed method can improve prediction performance for marked TPP models, and will likely work well in real case studies.

To summarize the simulations studies, we have shown throughout the section including multiple data configurations that the application of our proposed reweighting scheme can consistently and robustly improve the prediction performance of the common PP models utilizing the hidden variables. 

% \woody{Consider moving these intensity plot to the appendix.}
% \begin{figure}[!t]
% \centering
% \begin{subfigure}[b]{.32\linewidth}
% \includegraphics[width=\linewidth]{imgs/MAE_1.png}
% \caption{Syn Exp No.1}
% \end{subfigure}
% \hfill
% \begin{subfigure}[b]{.32\linewidth}
% \includegraphics[width=\linewidth]{imgs/MAE_1.png}
% \caption{Syn Exp No.2}
% \end{subfigure}
% \hfill
% \begin{subfigure}[b]{.32\linewidth}
% \includegraphics[width=\linewidth]{imgs/MAE_1.png}
% \caption{Syn Exp No.3}
% \end{subfigure}
% \begin{subfigure}[b]{.32\linewidth}
% \includegraphics[width=\linewidth]{imgs/MAE_4.png}
% \caption{Syn Exp No.4}
% \end{subfigure}
% \hfill
% \begin{subfigure}[b]{.32\linewidth}
% \includegraphics[width=\linewidth]{imgs/MAE_5.png}
% \caption{Syn Exp No.5}
% \end{subfigure}
% \hfill
% \begin{subfigure}[b]{.32\linewidth}
% \includegraphics[width=\linewidth]{imgs/MAE_6.png}
% \caption{Syn Exp No.6}
% \end{subfigure}
% \caption{Category-wise prediction MAE of synthetic experiments.}
% \label{fig:MAE_barchart}
% \end{figure}

\paragraph{Error Metrics}
Since in this study the conditional intensity functions for the sequence-generating models are specified and known, we directly compare the true conditional intensity $\lambda(t)$ against the fitted function $\hat{\lambda}(t)$ with parameters fitted by each method. Then the MAE across the time frame is calculated using a discrete grid of size $r$ in $t$, denoted by $\bar{t}_0,\bar{t}_2,\dots, \bar{t}_r$:
\begin{equation*}
\text{MAE} = \frac{1}{N_{\text{test}}} \sum_{idx = 1}^{N_{\text{test}}} \sum_{t=\bar{t}_0}^{\bar{t}_r} |\hat{\lambda}_{idx}(t)-\lambda_{idx}(t)|,
\end{equation*}
where $N_{\text{test}}$ denotes the number of testing event sequences. We take the average of errors across all testing sequences to obtain this error metric.

% Due to the limited space of the manuscript, the full error analysis of the numerical results for both the simulation study and the real case applications will be released in the online version instead.

\section{Details for Real Case Studies}
\label{sec:real-detail}
\paragraph{Netflix Movie Rating Data}
The data for each anonymous user include the date they rated a movie and the rating (from 1 to 5) the user provided for the corresponding movie. We append the movie genre information to each movie using the public website \url{IMDB.com}, and only look for positive ratings (3 to 5). We only select positive rating events from users because users tend to make fewer negative ratings as time goes by since they tend to focus more on movies they appreciate. This limits the use of event prediction for negative ratings. By doing this, each movie rating event for a single user is now characterized by time and the movie genre is denoted as a tuple $(t,m)$ where the genre is taken up as the mark variable. Furthermore, we divide the Netflix users into different categories based on their respective favorite movie genres given the ratings provided by them. We aim to predict the time and movie genre of the next rating event for ''new'' users not yet observed in the rating history data.

We take 50,000 entries by 300 Netflix users from the entire dataset to form the training dataset. The training set contains 300 user sequences scaled to a time frame of $[0,T)$ with $T=100$, containing movies from 45 genres. After pre-processing, the selected dataset contains 6712 discrete events from these 300 users over three years, then scaled to a time frame of $[0,T)$ with $T=100$. We fix the number of movie genres to be investigated to 45. In other words, the categorical mark has 45 potential values. These 300 users are divided into ten different clusters. For prediction, we further obtain the rating event sequences of 75 additional ''new'' users without obtaining their cluster information due to their lack of clear preference towards a certain genre. We set the dimension of the hidden embedding variable to $q=3$ and the number of bins $1/\delta=125$. Additionally, we set $\eta=5$.

For evaluation, since the true conditional intensity is no longer available, we now utilize different error metrics for performance comparison. For time prediction, we propose to calculate the MAE between the predicted time of the next event and the actual time for the testing event sequences. For mark prediction, we propose to calculate the accuracy of the top 5 predicted marks against the actual mark for the next event given the large number of potential mark values, abbreviated by Acc. In other words, the proportion of prediction cases where the top 5 guesses cover the true mark value of the upcoming event. Note the exponential Hawkes model is not capable of handling the mark variable, so it is excluded in this case.

\paragraph{Amazon Contact Data}
For Amazon, user contact intent prediction is crucial as it heavily contributes to the support system efficiency. When sellers need help from Amazon, such as how to create a listing, they often reach out to Amazon seller support through email, chat, or phone. As a common practice in customer support business, intents are assigned to the contacts for scalable support management. Though there is an abundance of literature in customer support contact intent prediction \citep{dong2021semi, Sarikaya2011}, there are a couple of challenges that few touch: ($i$) The support contact volume tends to vary significantly and unexpectedly, creating challenges for companies to prepare for the support capacity. Therefore, the ability to predict the time and intent of the next potential contact for customers becomes critical; ($ii$) With new sellers joining the platform every day, it can be challenging to provide confident and fair intent prediction given their relatively short history. One workaround is grouping sellers into clusters and predicting intents based on the cluster-level information. However, it is still challenging to keep the category information up to date. These two challenges indeed warrant the application of our proposed method, which strives to provide a prediction for user events without the knowledge of their respective cluster information.

The seller support contact dataset from Amazon presented in this work includes user contact events from thousands of sellers sampled from recent years \citep{dong2023ascisdata}. For each contact event, we have the information on the corresponding anonymous user ID, the time of the contact, and the identified contact intent marked by support associates. We regard the discrete intent as the mark variable, which can take any of the 117 possible values. Additionally, for data privacy purposes, the user identity is anonymized, and the contact time is polluted lightly with random noise.

We obtain from the data 3,000 user event sequences, where each sequence records the contact of an Amazon seller. Each event contains the time of the contact event and a discrete mark variable for the corresponding intent. For the mark variable, it can take any of the 117 possible values. The 3,000 users for training fall into 12 categories, and we further obtain 500 user event sequences without category information for testing. Our objective in this application is to predict the next contact time and intent for these 500 users in the testing dataset. For the counterfactual framework, we set $q=3$, the update period $\eta=5$, and the bin number to $1/\delta=125$. 

\end{document}